\documentclass[12pt,preprint]{aastex}
\usepackage{epsfig}

\newcommand{\msun}{{\rm M}_{\odot}}
\newcommand{\Hm}{\rm{H}^{-}}
\newcommand{\me}{\rm{e^{-}}}
\newcommand{\Hp}{\rm{H}^{+}}
\newcommand{\Dp}{\rm{D}^{+}}
\newcommand{\Hep}{\rm{He}^{+}}
\newcommand{\mH}{\rm{H}}
\newcommand{\mD}{\rm{D}}
\newcommand{\He}{\rm{He}}
\newcommand{\mHt}{\rm{H}_{2}}
\newcommand{\hd}{\rm{HD}}
\newcommand{\mHtp}{\rm{H}_{2}^{+}}
\newcommand{\htp}{\rm{H}_{3}^{+}}
\newcommand{\mC}{\rm{C}}
\newcommand{\cp}{\rm{C^{+}}}
\newcommand{\Cp}{\rm{C^{+}}}
\newcommand{\Cm}{{\rm C^{-}}}
\newcommand{\mO}{\rm{O}}
\newcommand{\Om}{\rm{O^{-}}}
\newcommand{\op}{\rm{O}^{+}}
\newcommand{\Op}{\rm{O}^{+}}
\newcommand{\co}{{\rm CO}}
\newcommand{\oh}{{\rm OH}}
\newcommand{\hto}{{\rm H_{2}O}}
\newcommand{\ch}{{\rm CH}}
\newcommand{\msi}{\rm{Si}}
\newcommand{\mSi}{\rm{Si}}
\newcommand{\sip}{\rm{Si^{+}}}
\newcommand{\sipp}{\rm{Si^{++}}}
\newcommand{\tmpt}[1]{\left(\frac{T}{300}\right)^{#1}}
\newcommand{\tmptscl}[2]{\left(\frac{T}{#1}\right)^{#2}}
\newcommand{\expf}[3]{\exp \left(#1\frac{#2}{#3}\right)}
\def\simless{\mathbin{\lower 3pt\hbox
   {$\rlap{\raise 5pt\hbox{$\char'074$}}\mathchar"7218$}}}   
\def\simgreat{\mathbin{\lower 3pt\hbox  
   {$\rlap{\raise 5pt\hbox{$\char'076$}}\mathchar"7218$}}} 

\shorttitle{Fragmentation of Cold Gas at Low Metallicity}
\shortauthors{Jappsen et al.}

\begin{document}
\title{Star Formation at Very Low Metallicity. IV. Fragmentation Does Not Depend on Metallicity for Cold Initial Conditions.}

\author{Anne-Katharina Jappsen\altaffilmark{1}}
\affil{School of Physics and Astronomy, Cardiff University, Cardiff, UK}
\email{jappsena@cardiff.ac.uk}

\author{Ralf S. Klessen}
\affil{Institut f\"ur Theoretische Astrophysik, Zentrum f\"ur
  Astronomie der Universit\"at Heidelberg, Heidelberg, Germany} 
\email{rklessen@ita.uni-heidelberg.de}

\author{Simon C. O. Glover\altaffilmark{2}}
\affil{Institut f\"ur Theoretische Astrophysik, Zentrum f\"ur
  Astronomie der Universit\"at Heidelberg, Heidelberg, Germany}
\email{sglover@ita.uni-heidelberg.de}
\and
\author{Mordecai-Mark Mac Low\altaffilmark{3}}
\affil{American Museum of Natural History, New York, NY, USA}
\email{mordecai@amnh.org}

\altaffiltext{1}{Formerly at: Canadian Institute for Theoretical Astrophysics, Toronto, ON,
  Canada}
\altaffiltext{2}{Formerly at: Astrophysikalisches Institut Potsdam, Potsdam, Germany}
\altaffiltext{3}{Also at the Max-Planck-Institut f\"ur Astronomie and
  the Institut f\"ur Theoretische Astrophysik, Zentrum f\"ur Astronomie
  der Universit\"at Heidelberg, Heidelberg, Germany}

\begin{abstract}
Primordial star formation appears to result in stars at least an order
of magnitude more massive than modern star formation. It has been proposed that the transition from primordial to modern initial
mass functions occurs due to the onset of effective metal line cooling
at a metallicity ${\rm Z}/{\rm Z}_{\odot} = 10^{-3.5}$.  However, these
simulations neglected molecular hydrogen cooling.  We perform simulations
using the same initial conditions, but including molecular cooling,
using a complex network that follows molecular hydrogen formation and also directly follows carbon monoxide and water.  We find that molecular hydrogen cooling allows roughly equivalent fragmentation to
proceed even at zero metallicity for these initial
conditions. The apparent transition just represents the point where
metal line cooling becomes more important than molecular cooling. In
all cases, the fragments are massive enough to be consistent with
models of primordial stellar masses, suggesting that the transition to
the modern initial mass function may be determined by other physics
such as dust formation. We conclude that such additional cooling
mechanisms, combined with the exact initial conditions produced by
cosmological collapse are likely more important than metal line
cooling in determining the initial mass function, and thus that there
is unlikely to be a sharp transition in the initial mass function at
${\rm Z}/{\rm Z}_{\odot} = 10^{-3.5}$.
\end{abstract}


\section{Introduction}

The formation of stars is a key process in the early universe with far-reaching consequences for cosmic reionization and galaxy formation \cite[][]{lb01,bl04,glover05}. The physical processes that govern stellar birth in a metal-free or metal-poor environment, however, are still very poorly understood. Numerical simulations of the thermal and dynamical evolution of gas in primordial protogalactic halos indicate that the metal-free first stars, the so called Population III, are expected to be very massive, with masses 
anywhere in the range 20--$2000\,{\msun}$ \cite[][]{abn02,bcl02,yoha06,oshn07}. In contrast, the stellar mass spectrum in the present-day universe is dominated by low-mass stars, peaking at around $0.2\,{\msun}$ \cite[][]{scalo98,kroupa02,chabrier03}. This means that at some stage of cosmic evolution there must have been a transition from primordial, high-mass star formation to the ``normal'' mode of star formation that dominates today. 
The discovery of extremely metal-poor subgiant stars in the Galactic halo with masses below one solar mass \cite[][]{christlieb02,bc05} indicates that this transition occurs at abundances considerably smaller  than the solar value. At the extreme end, these stars have iron abundances less than $10^{-5}$
times the solar value,  and carbon or oxygen abundances that are still $ \simless \,10^{-3}$ times the solar value. These stars are thus strongly  iron deficient, which  could be due to unusual abundance patterns produced by substantial mixing and fallback of ejecta during supernovae explosions of stars with masses between 20-130 M$_{\odot}$ \citep{ume03} or due to mass transfer from a close binary companion \citep{ry05,kom07}. There are hints for an increasing binary fraction with decreasing metallicity for these stars \cite[][]{lucatello05}. 

In a seminal paper, \citet{bfcl01} proposed that the transition from
high to low mass star formation should occur at a critical metallicity
Z$_{\rm crit} \approx 10^{-3.5}\,$Z$_{\odot}$ as a result of atomic
fine-structure line cooling from alpha elements, such as carbon and
oxygen, becoming important.  Their study is based on smoothed particle
hydrodynamics simulations of the collapse of cold gas in a top-hat
potential at densities below $n = 10^6\,$cm$^{-3}$. It assumes that
cooling from molecular hydrogen is negligible in the density and metallicity range considered. The  value of Z$_{\rm crit}$ can also be estimated analytically by calculating the metallicity required to produce a cooling rate equal to the rate of adiabatic compression heating for given halo properties \cite[][]{bl03,san06,fjb07}.  
However, \citet{bfcl01} noted that neglecting H$_2$ cooling could significantly influence the resulting fragmentation pattern. Along the same lines,  \citet{omu05} argued, based on the results of detailed 
one-zone calculations, that molecular cooling would indeed dominate the cooling over many orders 
of magnitude in density. They argued that dust-induced cooling at densities above $n \approx 10^{13}\,$cm$^{-3}$ may be a more viable mechanism to explain fragmentation in very low metallicity gas, a conjecture that has recently been confirmed by \citet{cgk07}.

The effects of molecular cooling at densities up to $n\approx 500\,$cm$^{-3}$ have been discussed 
by \citet[][hereafter Paper II]{jgkm07} in  three-dimensional collapse simulations of warm ionized gas in protogalactic halos for a wide range of environmental conditions.  This study used a time-dependent  chemical network running alongside the hydrodynamic evolution as described in 
\citet[][hereafter Paper I]{gj07}. The physical motivation was to investigate  whether small protogalaxies that formed within the relic H{\sc ii} regions left by neighboring Population III stars  could form subsequent generations of stars themselves, or whether the elevated temperatures and fractional ionizations found in these regions suppressed star formation until larger protogalaxies formed. 
We  found in Paper II that molecular hydrogen dominates the cooling of the gas for abundances up to at
least $10^{-2}$ Z$_{\odot}$. 
In addition, there was no evidence for fragmentation at densities below $500\,$cm$^{-3}$. 

To understand whether this apparent discrepancy with the \citet{bfcl01} results comes from including  molecular hydrogen cooling  or from adopting vastly different initial conditions --- here  ionized halo gas in a dark matter halo with a NFW density profile \cite[][]{nfw97}, there cold atomic gas in a top-hat halo --- we apply our time-dependent chemical network with molecular hydrogen cooling to \citet{bfcl01}-type initial conditions. Our study is structured as follows. In the next section, \S\ref{sec:method}, we give the details on the numerical set-up and the adopted initial conditions which we summarize in Table~\ref{tab:runs}. In \S\ref{sec:result} we report our results, and we summarize in \S\ref{sec:summary}.
 
\section{Numerical Set-Up}
\label{sec:method}
To adequately describe star formation in the early universe, it is necessary to follow the cooling and possible fragmentation of  the gas in the central regions of dark-matter halos 
over many orders of magnitude in density. Due to the nonlinear nature of the flow,  it is not known in
advance where and when gravitational collapse will occur, and consequently stars will form. To
compute the time evolution of the system we therefore resort to smoothed particle
hydrodynamics (SPH), which is a Lagrangian scheme to solve the equations of
hydrodynamics \cite[][]{ben90,mon92,mon05} coupled with the a simplified version of the
time-dependent  network discussed 
in \citet[][hereafter, Paper III]{glo07}. The fluid is represented by an ensemble of particles, 
and flow quantities are obtained by averaging over an appropriate subset of SPH particles. 
The method is able to resolve high density contrasts as particles are free to move, and so the 
particle concentration increases naturally in high-density regions. We use a modified version 
of the parallel SPH code GADGET \citep{spr05}.   Once the central parts of a collapsing 
fragment exceed a density $n = 10^5$ cm$^{-3}$ we introduce a ``sink particle'' \cite[][]{BAT95,JAP05}, which is
 able to accrete gas from its surroundings while keeping track of the mass and
 linear and angular momentum of the infalling material. Replacing collapsing high-density cores
 by accreting sink particles allows us to follow the dynamic evolution of the system
 over many local free-fall timescales.

\subsection{Chemistry and Cooling}
The chemical model used in our simulations is based on the detailed model of
low-metallicity gas chemistry outlined in Paper III. This model
includes cooling both from atomic fine structures lines and also
from the molecules $\mHt$, HD, CO, OH and ${\rm H_{2}O}$. It was
designed to accurately follow the formation and destruction of these 
molecular species over a wide range of temperatures and densities.

In Paper I,
we showed that the neglect of molecular species including metals can be
justified in simulations of gravitationally collapsing protogalactic
gas provided that $n < 500 / t_{\rm char} \: {\rm cm^{-3}}$, where
where $t_{\rm char}$ is the characteristic physical
timescale of interest, in Myr. At higher gas densities, molecule
formation from metals 
may become more significant, although its ultimate importance depends
on a number of factors, such as the temperature and metallicity of the
gas, and the presence (or absence) of an ultraviolet radiation
field. 

Although our chemical model is capable of accounting for the effects of dust, 
cosmic rays, and photodissociation by an external radiation field, we omit these complications in the present study.

A full list of the chemical reactions included in the chemical 
network used in this paper is given in Table~\ref{tab:chem_gas}, 
along with the corresponding rate coefficients and references to the source of 
the data. Given these rate coefficients, calculation of the appropriate chemical rates
is for the most part straightforward.
One minor exception is the case of $\mHt$ collisional
dissociation, which has a rate coefficient that depends on both temperature and
density. For the collisional dissociation of $\mHt$ by $\mH$ or $\mHt$ (reactions 9 \& 10), 
we calculate the rate coefficients using a function of the form
\begin{equation}
\log k_{\rm i}  = \left( \frac{n/n_{\rm cr}}{1 + n/n_{\rm cr}} \right)
\log k_{\rm i, h} + \left(\frac{1}{1 + n/n_{\rm cr}} \right) \log k_{\rm i, l},
\end{equation}
where $k_{\rm i, l}$ and $k_{\rm i, h}$ are the low density and high density
limits for the collisional dissociation rate due to collisions with species $i$ (which can
be found in Table~\ref{tab:chem_gas}), and where the critical density, $n_{\rm cr}$, 
is given by
\begin{equation}
\frac{1}{n_{\rm cr}} = \frac{x_{\mH}}{n_{\rm cr, \mH}} + 
\frac{x_{\mHt}}{n_{\rm cr, \mHt}},
\end{equation}
where $n_{\rm cr, \mH}$ and $n_{\rm cr, \mHt}$ are the critical densities in pure
atomic gas with an infinitesimally dilute quantity of $\mHt$ and in pure molecular 
gas respectively. Temperature dependent fits  for $n_{\rm cr, \mH}$ and 
$n_{\rm cr, \mHt}$ can be found in Table~\ref{tab:chem_gas}. For the collisional
dissociation of $\mHt$ by He (reaction 75), we use a slightly different expression,
taken from \citet{drcm87}:
\begin{equation}
\log k_{\rm He} = \log k_{\rm He, h} - \frac{\left(\log k_{\rm He, h} - 
\log k_{\rm He, l}\right)}{1.0 + (n / n_{\rm cr, He})^{1.09}}.
\end{equation}
Values for $k_{\rm He, l}$, $k_{\rm He, h}$ and $n_{\rm cr, He}$ are given in 
Table~\ref{tab:chem_gas}.

A referenced list of the thermal processes included in our models is given
in Table~\ref{tab:cool}. To compute the CO and ${\rm H_{2}O}$ rotational and vibrational
cooling rates, we use the multi-parameter fitting functions given in \citet{nk93}
and \citet{nlm95}. One of the parameters required is the effective column 
density  $\tilde{N}(M)$ of each coolant $M$ (where $M = {\rm CO}$ or 
${\rm H_{2}O}$). We calculate this in our simulations using the expression 
\citep{nk93}
\begin{equation}
\tilde{N}(M) = \frac{n(M)}{|{\bf{\nabla \cdot v}}|},
\end{equation}
where $n(M)$ is the local number density of species $M$. The \citet{nk93} and \citet{nlm95} 
treatments assume that only collisions with $\mHt$ are important in determining the CO 
or ${\rm H_{2}O}$ rotational cooling rates, as is appropriate in a fully molecular gas. 
However, in metal-poor gas, the $\mHt$ abundance will often be much smaller than the
abundances of atomic hydrogen or atomic helium. We therefore use an effective number
density $n_{\rm eff}$, given by \citep{pav02,sr03}
\begin{equation}
n_{\rm eff} = n_{\mHt} + \sqrt{2} n_{\mH} + 0.5 n_{\He}
\end{equation}
in place of the $\mHt$ number density in the \citet{nk93} and \citet{nlm95} fitting formulae.

Radiative cooling alone is unable to decrease the temperature of the gas below the
cosmic microwave background (CMB) temperature, on simple thermodynamical grounds. We account for this effect in
an approximate fashion by using a modified radiative cooling rate in our treatment of
the gas thermal energy equation. This modified rate, $\Lambda_{\rm mod}(T)$, is given
by
\begin{equation}
\Lambda_{\rm mod}(T) = \Lambda(T) - \Lambda(T_{\rm CMB}),
\end{equation}
where $\Lambda(T)$ and $ \Lambda(T_{\rm CMB})$ are the unmodified radiative 
cooling rates at $T$ and $T_{\rm CMB}$ respectively.

For additional details of our chemical networks and cooling function, 
please consult Papers I and III; further details of their implementation within Gadget 
are also given in Paper II and in \citet[][hereafter Paper V]{jmgk07}.

\subsection{Initial Conditions}
We model the initial dark matter distribution, following \citet{bfcl01}, 
as an isolated and roughly spherical overdensity described by the
top-hat approximation with small-scale fluctuations with an initial variance of $\sigma^2 \approx 0.1$ 
added. We consider a halo virializing at $z_{\rm vir} = 30$, which
corresponds to a roughly $3 \sigma$ peak with a total mass of $2
\times 10^6\, \msun$, corresponding to $2 \times 10^5\, \msun$ in
baryons. The overdense region has a proper radius of $150$ pc. For
more details on the set-up of the dark matter and the gas distribution
see \citet{bfcl01}.

We carry out numerical experiments with five
different metallicities ranging from ${\rm Z} = 0$ to
${\rm Z} = 0.1 \, {\rm Z}_{\odot}$. Aside from the difference in metallicity,
these calculations have identical initial conditions. The simulations
were initialized at a redshift $z = 100$, with an initial gas
temperature $T_{\rm gas, i}=200\, {\rm K}$. 
  To study the effect of resolution we ran simulations with three
  different resolutions. Our low resolution simulations, have 
$N_{\rm DM} = 13400$ dark matter particles and  $N_{\rm gas} = 65536$
gas particles. In these simulations, our numerical resolution of $M_{\rm res} = 150\,\msun$ 
was the same as in \citet{bfcl01}, 
Our higher resolution simulations with  $N_{\rm DM} = 107177$ and $N_{\rm gas} =
8 \times 10^{5}$ have
mass resolution $M_{\rm res} = 12.5\,\msun$. 
Finally, our highest resolution simulation has $N_{\rm DM} = 4057344$ and $N_{\rm gas} = 3.2 \times 10^{6}$, so its mass resolution is
$M_{\rm res} = 3.1\,\msun$. In all of our simulations,
both the dark matter and the gas particles were endowed with the same Hubble expansion and were 
set in rigid rotation with a spin parameter of $0.05$, just as in \citet{bfcl01}. 

We also consider a halo viralizing at a lower redshift of $z_{\rm vir} = 25$, with a total mass of $2\times 10^6\, \msun$, where the overdense region has a proper radius of 200~pc.   
We summarize the properties of all of the runs in Table~\ref{tab:runs}, where the runs with halos that collapse at a lower redshift are denoted with ''-C''. 

\section{Results}
\label{sec:result}
In each of our calculations the halo contracts rapidly and the gas density increases by several orders of magnitude. A disk-like structure builds up in the very center of each halo with a density of $n \approx 10^{3}\,$cm$^{-3}$ and a diameter of  10 to 20$\,$pc, as illustrated in Figure \ref{fig:proj}. 
The disk mass at the time illustrated is of order $M_{\rm disk} \sim
2$--$3 \times 10^4 \: \msun$, and so it is well-resolved even in the low resolution simulations.
The disk is supported by a combination of thermal pressure and rotation and its growth is fed by material falling in along filaments and sheets. This complex disk provides the background for the small-scale initial density fluctuations to grow, some of which become gravitationally unstable and collapse to form stars. 
In contrast to the results of \citet{bfcl01}, we find that fragmentation of the disk occurs in all of our
simulations, and that there is no evidence for a sharp transition in the initial mass function at ${\rm Z}
= 10^{-3.5} \: {\rm Z_{\odot}}$. 

\begin{figure}
\centering
\epsscale{0.8}
\plotone{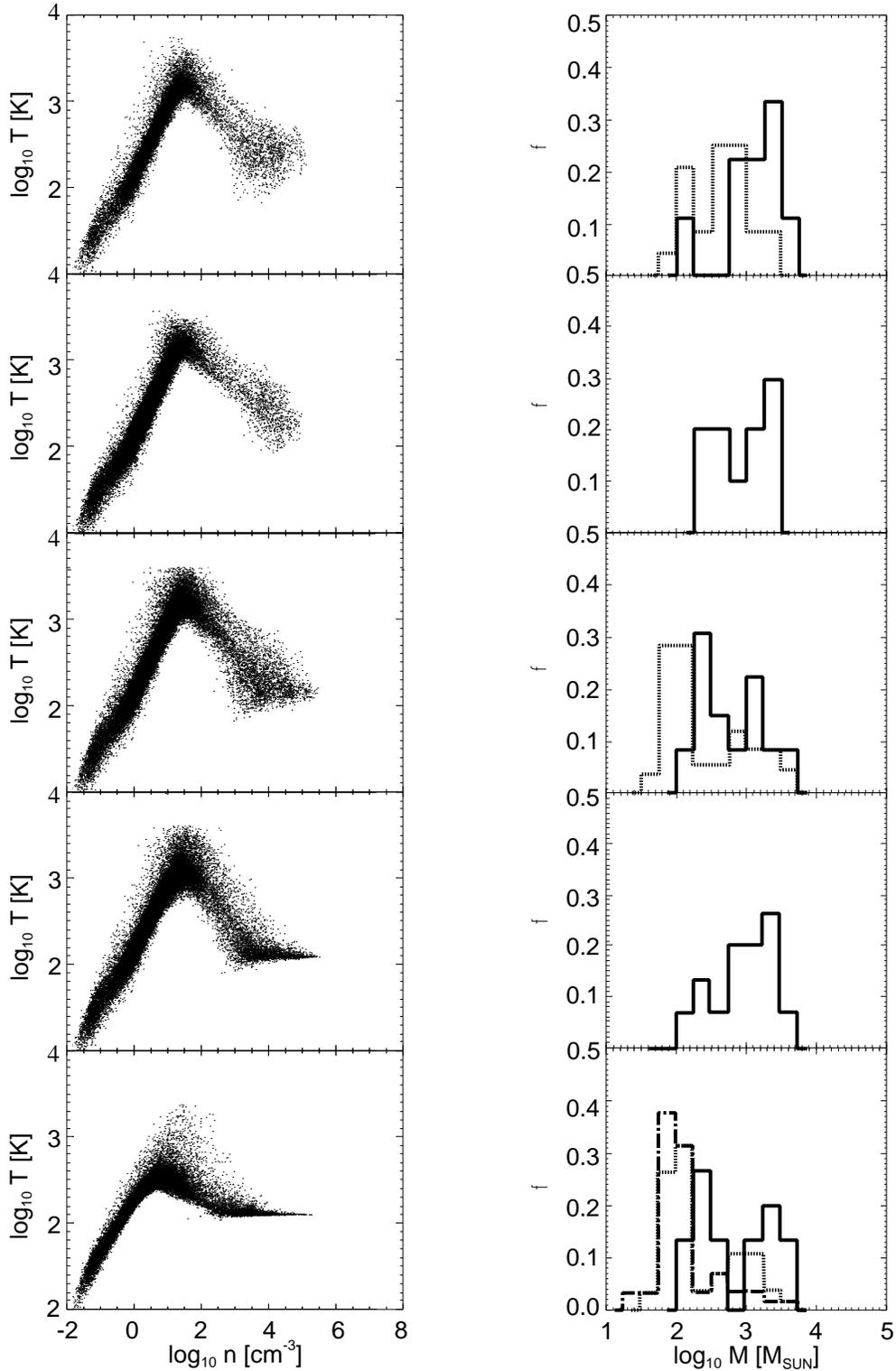}
\caption{{\it From top to bottom}: Gas properties and mass distribution of clumps at $z = 37.5$
for runs Z0, Z-4, Z-3, Z-2, and Z-1, arranged in order of increasing metallicity.
{\it Left}: Temperature versus number density of the gas.  {\it Right}: Mass distribution of the clumps represented by sink particles. In the panels for runs Z0, Z-3, and Z-1 we also show the results for 
the higher resolution runs Z0-BIG, Z-3-BIG, and Z-1-BIG ({\it dotted lines}). The {\it dot-dashed line} in the bottom panel represents the run with the highest resolution Z-1-BIG2. }
\label{fig:t-rho}
\end{figure}

This result can be understood as follows. Regardless of the metal
content, H$_2$ line cooling maintains
the gas temperature at densities above $10^3\,$cm$^{-3}$ close to the CMB temperature ($T_{\rm CMB} = 105\,$K at the
redshift $z=37.5$ considered in Figure~\ref{fig:t-rho}); even in the
primordial case, the gas temperature is within a factor of two of this
value. The resulting equation of state is therefore approximately
isothermal (at high densities) or softer than isothermal (at low
densities). The thermal behavior of the gas is depicted in the left
column of Figure \ref{fig:t-rho}. 
 
\begin{figure}
\centering
\epsscale{0.8}
\plotone{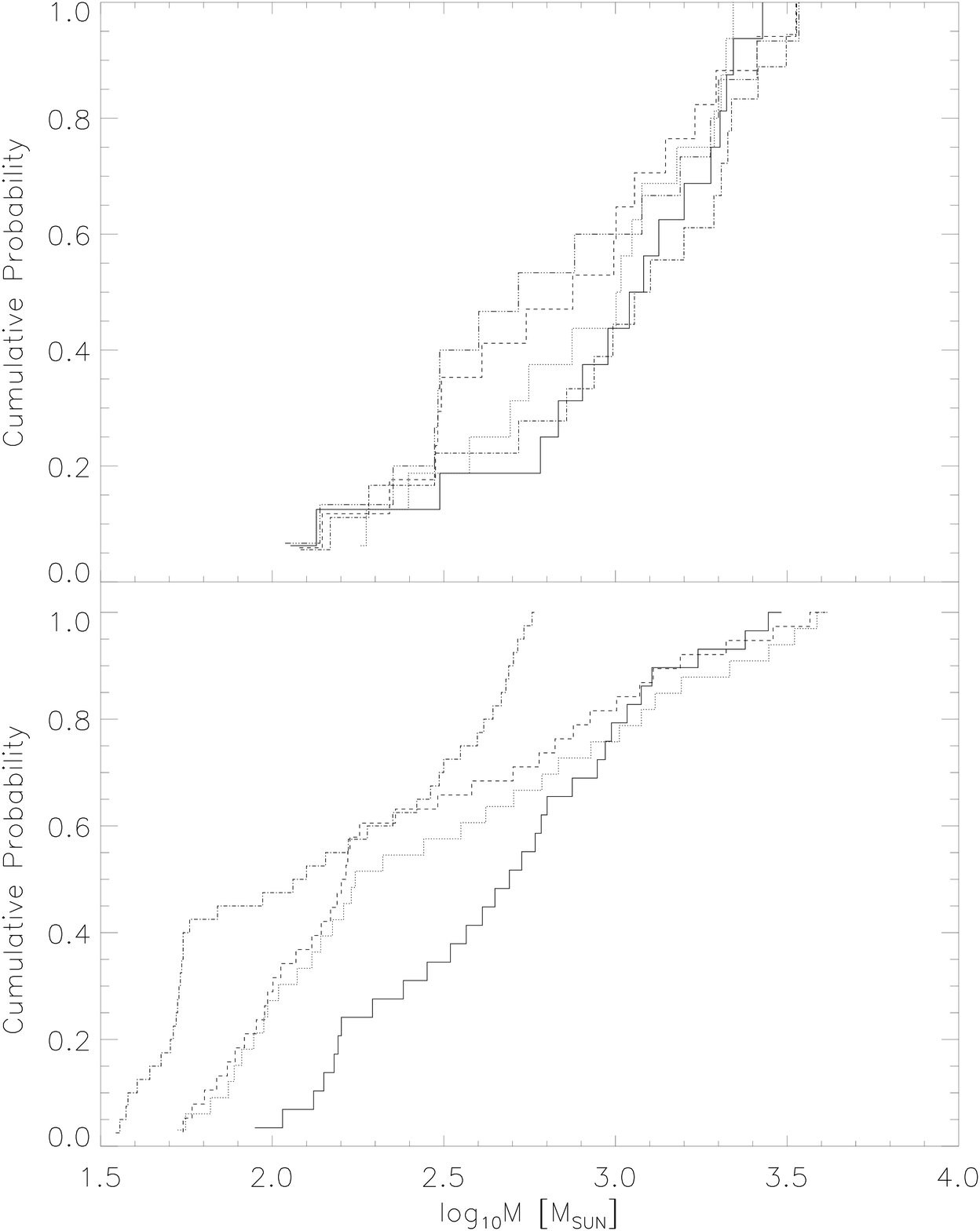}
\caption{{\it Top panel}: Cumulative distribution function of the clump masses for the low resolution runs Z0 ({\it solid line}), Z-4 ({\it dotted line}), Z-3 ({\it dashed line}), Z-2 ({\it dot-dashed line}), and Z-1 ({\it dot-dot-dashed line}).
{\it Bottom panel}: Cumulative distribution function of the clump masses for the higher resolution runs Z0-BIG ({\it solid line}), Z-3-BIG ({\it dotted line}), Z-1-BIG ({\it dashed line}) and Z-1-BIG-C ({\it dot-dashed line}).}
\label{fig:ks}
\end{figure}

The  physical conditions where the equation of state changes from softer than isothermal to isothermal or stiffer, i.e. from a regime where the gas cools with increasing density to one where the temperature remains constant or rises again, imprint a characteristic mass in the fragmentation process \cite[][]{JAP05, larson2005}.
Because this fragmentation occurs at approximately the same density and temperature in each 
simulation, the mass distribution of the fragments does not vary greatly from simulation to 
simulation. In each of our low resolution simulations, between 16 and 18 fragments are formed,
with masses in the range $10^2\,$M$_{\odot} \le M \le 5 \times 10^3\,$M$_{\odot}$. The resulting mass 
spectra are shown as the solid lines in the right-hand column of Figure~\ref{fig:t-rho}. For 
comparison, the thermal Jeans mass for gas with temperature $T_{\rm
  gas} = T_{\rm CMB}$ is approximately $2.3 \times 10^3 \: {\rm
  M}_{\odot}$ at $n = 10^3 \: {\rm cm^{-3}}$,
  the density above which H$_2$ line cooling maintains the gas
as approximately isothermal,
and  $230 \: {\rm M_{\odot}}$ at  $n = 10^{5} \: {\rm cm^{-3}}$, the
 density above which we form sinks.  
To quantify whether the mass spectra differ significantly, we can use the two-sided 
Kolmogorov-Smirnov test to determine for each of our low resolution, metal-enriched simulations 
the probability that the mass spectrum produced is drawn from the same underlying distribution
as in the run Z0. We show the corresponding cumulative distribution functions of the clump masses in the top panel of Figure~\ref{fig:ks}.  We find probabilities of 0.63, 0.22, 0.99 and 0.11 for runs Z-4, Z-3, Z-2 and Z-1 respectively, indicating that the differences between the fragment mass spectra are in general 
not statistically significant. We also determine the probabilities for the runs Z-3, Z-2 and Z-1 that the mass spectrum is drawn from the same underlying distribution as in the Z-4 simulation. The probabilities are 0.73, 0.63, 0.5, respectively. This shows that there is no significant change in the clump mass function at ${\rm Z} = 10^{-3.5} \: {\rm Z_{\odot}}$.

In our high resolution simulations, we find a greater degree of fragmentation and a
broad fragment mass distribution. However, just as in our low resolution simulations,
the mass distribution is dominated by fragments with masses $M > 100 \: \msun$.
We find very few objects with masses close to our resolution limit of $12.5 \: \msun$,
and there is no indication from our current results that we are missing a substantial
population of objects with masses below this limit. As before, we can compare the
fragment mass distributions resulting from our three high-resolution simulations with
the K-S test. We show our results in the bottom panel of Figure~\ref{fig:ks}. We find that the probabilities that the fragment mass distributions in runs
Z-3-BIG and Z-1-BIG are drawn from the same underlying distribution as in run Z0-BIG
are 0.11 and 0.02 respectively, indicating that there is a statistically significant difference. If we compare the runs Z-3-BIG and Z-1-BIG then we determine a probability of 0.93 that their drawn from the same underlying distribution. The run with zero metallicity results in fragments with masses above $100 \: {\rm M_{\odot}}$, whereas the runs with metallicity also produce fragments with masses between 50 and $100 \: {\rm M_{\odot}}$. Although we see some slight influence from the metal-line cooling, we see
no evidence for any low-mass star formation: all of the clumps formed are well-resolved, and have
masses consistent with recent determinations of the plausible mass range for Population III stars
(see e.g.\ \citealt{oshn07}). We see no indication that metal-line cooling at these temperatures and
densities is capable of producing low-mass stars.

We note that the maximum gas temperature during collapse depends on the metallicity.
In the zero-metallicity case, the peak temperature $T_{\rm peak}$ 
reached due to compressional heating during the initial contraction of the gas is $\sim 6 \times 10^3\,$K. 
 As the metallicity increases, however, $T_{\rm peak}$ decreases. In particular,
the run with ${\rm Z} = 0.1 \: {\rm Z}_{\odot}$ shows a much smaller peak temperature of $\sim 10^3\,$K. This decrease in $T_{\rm peak}$ reflects the growing importance of metal-line cooling in cold gas as the metallicity increases. From Figure \ref{fig:t-rho} it is apparent that the metals begin to affect $T_{\rm peak}$ once the metallicity exceeds ${\rm Z} = 10^{-3} \: {\rm Z}_{\odot}$. This is a lower threshold
than found in \citet{jgkm07}, but the difference is easily understood as a consequence of the fact 
that in these simulations we start with cold gas, and that the H$_2$ cooling rate of the cold, 
low-density gas in these simulations is much lower than  H$_2$ cooling rate in the warm, 
low-density gas studied in \citet{jgkm07}. Our threshold of ${\rm Z} \sim 10^{-3} \: {\rm Z}_{\odot}$ is in 
reasonable agreement with the \citet{bfcl01} threshold of ${\rm Z} \sim 10^{-3.5} \: {\rm Z}_{\odot}$; however, 
the key difference here is that at lower metallicities, the H$_2$ alone is sufficient to cool the gas, 
and the outcome of the fragmentation process has very little dependence on whether the cooling 
comes from the H$_2$ or from the metals.

We also test the hypothesis that the fragment mass spectra of run Z-1-BIG and of the higher resolution run Z-1-BIG2 are drawn from the same distribution and find a probability of 0.98. Since the number of fragments are also quite similar we conclude that our runs Z-1-BIG, Z-3-BIG and Z0-BIG are converged.

\begin{figure}
\centering
\epsscale{0.9}
\plotone{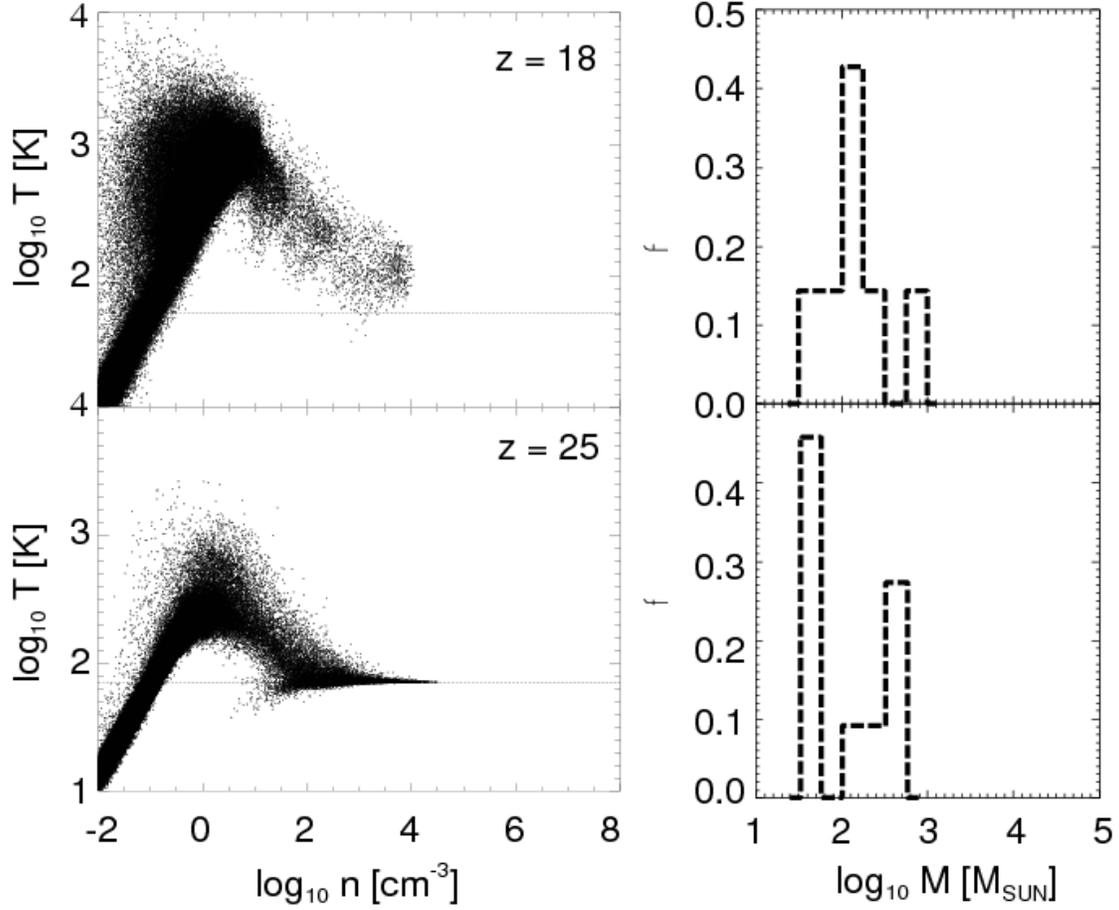}
\caption{{\it Top panel}: Gas properties and mass distribution of clumps at $z = 18$ for the run Z0-BIG-C. In this run no clumps have formed yet at $z = 25$ when the halo starts to collapse. {\it Bottom panel}: Gas properties and mass distribution of clumps at $z = 25$ for the run Z-1-BIG-C. The {\it dashed line} shows the CMB temperature.}
\label{fig:lowz}
\end{figure}

In all of our 
simulations the metal-enriched gas is able to cool to the CMB
temperature of approximately 110~K at this redshift (Figure~\ref{fig:t-rho}). 
This is also true to within a factor of 2 for the primordial case. H$_2$ can act
as a coolant for the gas down to a temperature of approximately 200~K,
which is close to the CMB temperature at this redshift. Since this
could be seen as a reason for the lack of
influence of the metals that we find, we perform two runs Z0-BIG-C and Z-1-BIG-C with
halos that collapse at a lower redshift $z = 25$. 
In Figure~\ref{fig:lowz} we show the results of these two runs. At a
redshift $z=$ 25 the CMB temperature is approximately 71~K and the run
Z-1-BIG-C is able to cool down to the CMB temperature due to the
combination of efficient metal-line cooling and H$_2$ cooling. In the
primordial case however the gas is not able to cool at all at this
redshift. 
However, by redshift $z=18$, a substantial amount of the gas is
nevertheless able to cool, collapse, and fragment from H$_2$ cooling, even though H$_2$ cannot cool much
 below $\sim 200$~K, and so cannot cool down to the CMB temperature at this redshift.
We compare the fragment mass distributions resulting from the primordial and the metal-enriched case with the K-S test. We find that the probability that the fragment mass distributions in runs Z0-BIG-C and Z-1-BIG-C are drawn from the same underlying distribution is 0.11. The probability that runs Z-1-BIG-C and run Z-1-BIG are drawn from the same underlying distribution is 0.05 (see Figure~\ref{fig:ks}). The fragments of run Z-1-BIG-C have masses down to $30 \: {\rm M_{\odot}}$ whereas the fragments created in run Z-1-BIG have masses above $50 \: {\rm M_{\odot}}$. In both runs the gas is able to cool down to the CMB temperature due to the combination of metal-line cooling and H$_2$ cooling. Since the CMB temperature is smaller for the run Z-1-BIG-C, the gas can fragment to smaller masses. This demonstrates that in runs with higher metallicity, where the gas is able to cool down to the CMB temperature, this temperature determines the dynamical behavior (c.f.\ the similar result found by
\citealt{sm08}). Nevertheless, in all of the runs, the fragments have masses that are consistent with the expected masses of Population III stars, and there is no evidence
for the formation of low-mass Population II stars.     


\section{Summary and Conclusion}
\label{sec:summary}
The current study shows that gas in simulations of the type considered here, with low initial temperature,
moderate initial rotation, and a top-hat dark-matter overdensity, will readily fragment into multiple
objects with masses above $30 \: {\rm M_{\odot}}$, regardless of metallicity, provided that enough $\mHt$ is present to cool the gas.  Rotation
leads to the build-up of massive disk-like structures in these simulations which allow smaller-scale fluctuations to grow and become gravitationally unstable to form protostars. The resulting mass 
spectrum of fragments peaks at a few hundred solar masses, roughly
corresponding to the thermal  
Jeans mass in the disk-like structure.

Our major
conclusion is that the metallicity threshold at
${\rm Z} = 10^{-3.5} \: {\rm Z}_{\odot}$ reported by \citet{bfcl01} does not represent a critical metallicity above
which gas fragments and below which it does not. Rather, this threshold simply marks the point
at which metal-line cooling becomes more important than $\mHt$ cooling at the gas densities
and temperatures relevant for fragmentation in the disk. This metallicity threshold only represents
a critical metallicity for fragmentation in these simulations if $\mHt$ formation is strongly suppressed
or not considered \cite[e.g.][]{bfcl01}. This scenario may be relevant if the extragalactic UV background
is strong \citep{HAI00}, although recent work suggests that even in this case, suppression 
of $\mHt$ formation is less important than previously supposed \citep{wa07,os07b}.

We also find that the mass spectrum of fragments formed in these simulations
has very little dependence on the metallicity of the gas. 
In our low-resolution simulations, many of the fragments are
only marginally resolved, but we obtain very similar results from our high-resolution
simulations, in which the fragments are well-resolved, suggesting that this lack of 
influence is real.

We further note that simulations performed by \citet{ss07} that follow the evolution of cold, 
metal-enriched gas from cosmological initial conditions until collapse do find tentative 
evidence for a metallicity
threshold at around ${\rm Z} = 10^{-3} \: {\rm Z}_{\odot}$ that must be surmounted before fragmentation will 
occur. On the other hand, the simulations presented in Paper V that follow the collapse of hot, 
initially ionized gas into isolated NFW halos find no evidence for any fragmentation at or above 
this metallicity. We conclude that the question of whether there is a critical metallicity below which the formation of low-mass stars is impossible remains unresolved. Our current results show no evidence
for such a critical metallicity: fragmentation occurs even in our ${\rm Z} = 0$ simulation, due only to
efficient $\mHt$ cooling, and the mass spectrum of the fragments formed does not appear to
differ significantly from those obtained from our higher metallicity simulations. 

However, the
degree of fragmentation occurring in these simulations appears to be a consequence of the 
initial conditions chosen: simulations using
substantially different initial conditions find
very different results \citep{jmgk07,ss07}. Another example arises if
we compare our
zero metallicity results with the results of \citet{abn02}. These authors use proper cosmological initial conditions and in their simulation only one core forms. Thus the fragmentation in our simulation with primordial gas seems to be a generic outcome of the adopted initial conditions. It is therefore important to consider whether a metallicity threshold appears in simulations with different initial conditions and what the most realistic initial conditions are. To make further progress in understanding the role 
(if any) that metal-line cooling plays in promoting fragmentation, we need to develop a much better
understanding of how metals are initially dispersed into the high-redshift interstellar and
intergalactic medium, in order to be able to select the most appropriate initial conditions 
for our simulations.

Our results call attention to an alternative scenario for the transition to the present-day low-mass IMF.  \citet{omu05} stress the importance of dust-induced cooling at high densities, which  may become
important at metallicities as low as ${\rm Z} \approx 10^{-6} $ -- $10^{-5} \: {\rm Z}_{\odot}$ depending on the adopted dust model \citep{sch06}. Recent numerical simulations by \citet{cgk07} support this point of view and predict the existence of low-mass stars and even brown dwarfs with metal abundances  of $\sim 10^{-5}$ the solar value \cite[see also][]{to06}.

\begin{figure}
\centering
\epsscale{0.8}
\plotone{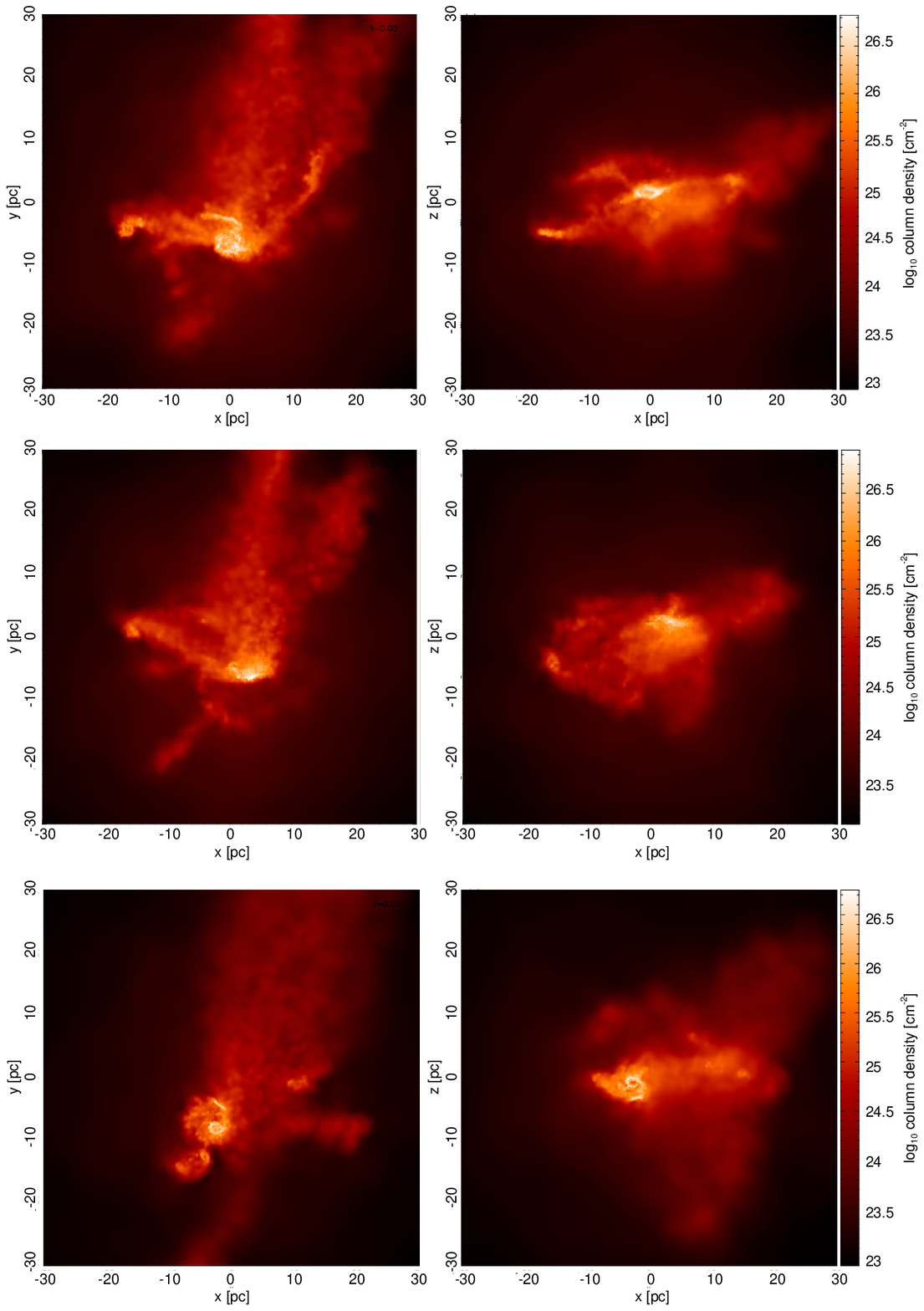}
\caption{{\it From top to bottom}: Projected gas number density in runs Z0, Z-3, and Z-1 at $z = 35$. {\it Left}: Face-on view. {\it Right}: Edge-on view. The size of the box is 60 pc.}
\label{fig:proj}
\end{figure}

\acknowledgments We acknowledge useful discussions with Tom Abel,
 Volker Bromm, Paul C.\ Clark, Kazuyuki Omukai, and Naoki Yoshida, and 
thank the anonymous referee for valuable feedback.   
This research was supported in part by the US National Science Foundation under Grants No.\ PHY05-51164, AST03-07854 and AST08-06558, 
and by the German Science Foundation (DFG) via the Emmy Noether Program under Grant No.\ KL1358/1 and via the Priority Program SFB 439 {\em Galaxies in the Early Universe}. AKJ acknowledges support by the
Human Resources and Mobility Programme of the European Community under
the contract MEIF-CT-2006-039569. Computations were performed on
the McKenzie cluster at the Canadian Institute for Theoretical
Astrophysics. RSK, SCOG and AKJ thank the staff and scholars of the Kavli
Institute for Theoretical Physics for their hospitality during the
final stages of the preparation of this paper, while M-MML thanks the Deutscher Akademischer Austausch Dienst and the Max-Planck-Gesellschaft for research stipends that helped support this work. Figure~\ref{fig:proj} was produced using SPLASH, a visualization package for SPH written by \citet{pri07}.

\clearpage

\clearpage

\begin{deluxetable}{llllrrcc}
\tablewidth{0pt}
\tablecaption{List of the runs discussed in this paper}
\tablehead{\colhead{Run} & \colhead{$Z$\tablenotemark{a}} & \colhead{$\lambda$\tablenotemark{b}} & \colhead{$N_{\rm gas}$\tablenotemark{c}} & \colhead{$N_{\rm clump}$\tablenotemark{d}}
\\\colhead{} &\colhead{($Z_{\odot}$)} & \colhead{} & \colhead{} & \colhead{}}
\startdata
Z0 & 0.0 & 0.05 & 65536 & 16\\ 
Z-4 & $10^{-4}$ & 0.05 & 65536 & 16\\
Z-3 & $10^{-3}$ & 0.05 & 65536 & 17\\
Z-2 & $10^{-2}$ & 0.05 & 65536 & 18\\
Z-1 & $10^{-1}$ & 0.05 & 65536 & 17\\
Z0-BIG & 0.0 & 0.05 & $8 \times 10^5$ & 29\\ 
Z-3-BIG & $10^{-3}$ & 0.05 & $8 \times 10^5$ & 33\\
Z-1-BIG & $10^{-1}$ & 0.05 & $8 \times 10^5$ & 38\\
Z-1-BIG2 & $10^{-1}$ & 0.05 & $3.2 \times 10^6$ & 38\\
Z0-BIG-C & 0.0 & 0.05 & $8 \times 10^5$ & 32\\
Z-1-BIG-C & $10^{-1}$ & 0.05 & $8 \times 10^5$ & 40\\ 
\enddata
\tablenotetext{a}{Metallicity of the gas.}
\tablenotetext{b}{Spin parameter of the halo.}
\tablenotetext{c}{Number of SPH particles representing the gas.}
\tablenotetext{d}{Number of clumps at $z = 18.0$ for the runs marked with "-C" and for all other runs at $z = 37.5$.}
\label{tab:runs}
\end{deluxetable}

\begin{deluxetable}{llllc}
\tablewidth{0pt}
\tabletypesize{\footnotesize}
\tablecaption{List of reactions in our chemical model. \label{tab:chem_gas}}
\tablehead{No.\  & Reaction & Rate coefficient $({\rm cm}^{3} \: {\rm s}^{-1})$ &
& Ref.\ } 
\startdata
& & & & \\
1 & $\mH  + \me  \rightarrow \Hm + \gamma$ & 
$k_{1} = {\rm dex}[-17.845 + 0.762 \log{T} + 0.1523 (\log{T})^{2}$ & & 1 \\
& & $\phantom{k_{1}= {\rm dex}[} \mbox{} - 0.03274 (\log{T})^{3}] $ &
$T \le 6000 \: {\rm K}$ & \\
& & $ \phantom{k_{1}} = {\rm dex}[-16.420 + 0.1998 (\log{T})^{2}$ & & \\ 
& & $ \phantom{k_{1} = {\rm dex}} \mbox{}-5.447  \times 10^{-3}  (\log{T})^{4}$ & & \\ 
& & $ \phantom{k_{1} = {\rm dex}} \mbox{}+ 4.0415 \times 10^{-5} (\log{T})^{6}]$ 
& $T > 6000 \: {\rm K}$ & \\
& & & & \\
2 & $\Hm  + \mH  \rightarrow \mHt + \me$ &
$k_{2} = 1.5 \times 10^{-9}$ & $T \le 300 \: {\rm K}$& 2 \\
& & $\phantom{k_{2}} = 4.0 \times 10^{-9} T^{-0.17}$ & $T > 300 \: {\rm K}$ & \\
& & & & \\
3 & $\mH  + \Hp  \rightarrow \mHtp + \gamma$ &
$k_{3} = {\rm dex}[-19.38 - 1.523 \log{T} $ & & 3 \\
& & $\phantom{k_{3}} \mbox{}+1.118 (\log{T})^{2}  - 0.1269 (\log{T})^{3}]$ & & \\
 & & & & \\
4 & $\mH + \mHtp \rightarrow \mHt + \Hp$ & $k_{4} = 6.4 \times 10^{-10}$ & & 4 \\
 & & & & \\
5 & $\Hm  + \Hp  \rightarrow \mH + \mH$ & 
$k_{5} = 5.7 \times 10^{-6} T^{-0.5}   +  6.3 \times 10^{-8} $ & & 5 \\
& & $\phantom{k_{5}} \mbox{} - 9.2 \times 10^{-11} T^{0.5}  +  4.4 \times 10^{-13} T$ & \\
& & & & \\
 6 & $\mHtp + \me \rightarrow \mH + \mH$ & $k_{6} = 
 1.0\times 10^{-8}$ &  $T \le 617 \: {\rm K}$ & 6 \\
 & & $\phantom{k_{6}} = 1.32 \times 10^{-6} T^{-0.76}$ & $T > 617 \: {\rm K}$ & \\
 & & & & \\
7 & $\mHt + \Hp  \rightarrow \mHtp + \mH$ &
$k_{7} = [- 3.3232183 \times 10^{-7}$ & & 7 \\
 & & $\phantom{k_{7}=}  \mbox{} + 3.3735382 \times 10^{-7}  \ln{T}$ & & \\
 & & $\phantom{k_{7}=}  \mbox{} - 1.4491368 \times 10^{-7}  (\ln{T})^2$ & & \\
 & & $\phantom{k_{7}=}  \mbox{} + 3.4172805 \times 10^{-8}  (\ln{T})^3$ & & \\
 & & $\phantom{k_{7}=}  \mbox{} - 4.7813720 \times 10^{-9}  (\ln{T})^4$ & & \\
 & & $\phantom{k_{7}=}  \mbox{} + 3.9731542 \times 10^{-10} (\ln{T})^5$ & & \\
 & & $\phantom{k_{7}=}  \mbox{}  - 1.8171411 \times 10^{-11}  (\ln{T})^6$ & & \\
 & & $\phantom{k_{7}=}  \mbox{}  + 3.5311932 \times 10^{-13} (\ln{T})^7 ]$ & & \\
 & & $\phantom{k_{7}=} \mbox{} \times \exp \left(\frac{-21237.15}{T} \right)$ & & \\
 & & & & \\
8 & $\mHt + \me  \rightarrow  \mH + \mH +  \me$ &
$ k_{8} = 3.73 \times 10^{-9} T^{0.1121} \exp\left(\frac{-99430}{T}\right) $
& & 8 \\
& & & & \\
9 & $\mHt + \mH  \rightarrow  \mH + \mH + \mH$  & 
$ k_{\rm 9, l} = 6.67 \times 10^{-12} T^{1/2} \exp \left[-(1+ \frac{63590}{T}) \right]$ & & 9 \\
& & & & \\
& & $k_{\rm 9, h} = 3.52 \times 10^{-9} \expf{-}{43900}{T}$ & & 10 \\
& & & & \\
& & $n_{\rm cr, H} = {\rm dex}\left[3.0 - 0.416 \log \left(\frac{T}{10000}\right) - 0.327  
\left\{\log \left(\frac{T}{10000}\right)\right\}^{2} \right]$ & & 10 \\
& & & & \\
10 & $\mHt + \mHt \rightarrow  \mHt + \mH + \mH$ & 
$k_{\rm 10, l} = \frac{5.996 \times 10^{-30} T^{4.1881}}{(1.0 + 6.761 \times 10^{-6} T)^{5.6881}}
 \exp \left(-\frac{54657.4}{T} \right)$ & & 11 \\
& & & & \\
& & $k_{\rm 10, h} = 1.3 \times 10^{-9} \expf{-}{53300}{T}$ & & 12 \\
& & & & \\
& & $n_{\rm cr, H_{2}} = {\rm dex}\left[4.845 - 1.3 \log \left(\frac{T}{10000}\right) + 1.62  
\left\{\log \left(\frac{T}{10000}\right)\right\}^{2} \right]$ & & 12 \\
& & & & \\
11 & $\mH  + \me  \rightarrow \Hp + \me + \me$ & 
$k_{11} =  \exp[-3.271396786 \times 10^{1}$ & & 13 \\
& & $\phantom{k_{11}=}  \mbox{}  + 1.35365560 \times 10^{1} \ln T_{\rm e}$ & & \\
& & $\phantom{k_{11}=}  \mbox{}  - 5.73932875 \times 10^{0} (\ln T_{\rm e})^{2}$ & & \\
& & $\phantom{k_{11}=}  \mbox{}  + 1.56315498 \times 10^{0} (\ln T_{\rm e})^{3}$ & & \\
& & $\phantom{k_{11}=}  \mbox{}  -  2.87705600 \times 10^{-1} (\ln T_{\rm e})^{4}$ & & \\
& & $\phantom{k_{11}=}  \mbox{}  + 3.48255977 \times 10^{-2} (\ln T_{\rm e})^{5}$ & & \\
& & $\phantom{k_{11}=}  \mbox{}   - 2.63197617 \times 10^{-3} (\ln T_{\rm e})^{6}$ & & \\ 
& & $\phantom{k_{11}=}  \mbox{}  + 1.11954395\times 10^{-4} (\ln T_{\rm e})^{7}$ & & \\
& & $\phantom{k_{11}=}  \mbox{}   -  2.03914985 \times 10^{-6} (\ln T_{\rm e})^{8}]$ & & \\
& & & & \\
12 & $\mD + \me \rightarrow \Dp + \me + \me$ & $k_{12} = k_{11}$ & & --- \\
& & & & \\
13 & $\Hp  + \me  \rightarrow \mH +  \gamma$ & 
 $k_{13, {\rm A}} = 1.269 \times 10^{-13} \left(\frac{315614}{T}\right)^{1.503}$ & Case A & 14 \\
 & & $\phantom{k_{13}=} \mbox{} \times
 [1.0+ \left(\frac{604625}{T}\right)^{0.470}]^{-1.923} $ & & \\
& & $k_{13, {\rm B}} = 2.753 \times 10^{-14} \left(\frac{315614}{T}\right)^{1.500}$ & Case B & 14 \\
 & & $\phantom{k_{13}=} \mbox{} \times
 [1.0+ \left(\frac{115188}{T}\right)^{0.407}]^{-2.242} $ & & \\
 & & & & \\
14 & $\Dp + \me \rightarrow \mD + \gamma$ & $k_{14} = k_{13}$ & & --- \\
 & & & & \\
15 & $\Hm  + \me  \rightarrow \mH + \me + \me$ &
$ k_{15}  = \exp [-1.801849334 \times 10^{1}$ & & 13 \\
& & $\phantom{k_{15}=}  \mbox{} + 2.36085220 \times 10^{0} \ln T_{\rm e}$ & & \\
& & $\phantom{k_{15}=} \mbox{} - 2.82744300 \times 10^{-1} (\ln T_{\rm e})^{2}$ & & \\
& & $\phantom{k_{15}=} \mbox{}  +1.62331664\times 10^{-2} (\ln T_{\rm e})^{3}$ & & \\
& & $\phantom{k_{15}=} \mbox{} -3.36501203 \times 10^{-2} (\ln T_{\rm e})^{4}$ & & \\
& & $\phantom{k_{15}=} \mbox{}   +1.17832978\times 10^{-2} (\ln T_{\rm e})^{5}$ & & \\ 
& & $\phantom{k_{15}=} \mbox{}  -1.65619470\times 10^{-3} (\ln T_{\rm e})^{6}$ & & \\ 
& & $\phantom{k_{15}=} \mbox{}   +1.06827520\times 10^{-4} (\ln T_{\rm e})^{7}$ & & \\
& & $\phantom{k_{15}=} \mbox{}  -2.63128581\times 10^{-6} (\ln T_{\rm e})^{8} ]$ & & \\
 & & & & \\
 & & & & \\
 & & & & \\
16 & $\Hm  + \mH  \rightarrow  \mH + \mH +  \me$ &
$k_{16} = 2.5634 \times 10^{-9} T_{\rm e}^{1.78186}$ & $ T_{\rm e} \le 0.1 \: \rm{eV}$ & 13 \\
& & $\phantom{k_{16}} = \exp[-2.0372609 \times 10^{1}$ & & \\
& & $\phantom{k_{16}=}  \mbox{}+1.13944933 \times 10^{0} \ln T_{\rm e}$ & & \\
& & $\phantom{k_{16}=} \mbox{}-1.4210135 \times 10^{-1} (\ln T_{\rm e})^{2}$ & & \\
& & $\phantom{k_{16}=} \mbox{}+8.4644554 \times 10^{-3} (\ln T_{\rm e})^{3}$ & & \\
& & $\phantom{k_{16}=}  \mbox{}-1.4327641 \times 10^{-3} (\ln T_{\rm e})^{4}$  & & \\
& & $\phantom{k_{16}=}  \mbox{}+2.0122503 \times 10^{-4} (\ln T_{\rm e})^{5}$ & & \\
& & $\phantom{k_{16}=}  \mbox{}+8.6639632 \times 10^{-5} (\ln T_{\rm e})^{6}$ & & \\
& & $\phantom{k_{16}=}  \mbox{}-2.5850097 \times 10^{-5} (\ln T_{\rm e})^{7}$ & & \\
& & $\phantom{k_{16}=} \mbox{}+ 2.4555012\times 10^{-6} (\ln T_{\rm e})^{8}$ & & \\
& & $\phantom{k_{16}=} \mbox{} -8.0683825\times 10^{-8} (\ln T_{\rm e})^{9}]$ & 
$T_{\rm e} > 0.1 \: \rm{eV}$ & \\ 
& & & & \\
17 & $\Hm + \Hp   \rightarrow \mHtp + \me$ & 
$k_{17}= 6.9\times 10^{-9}  T^{-0.35}$  & $T \le 8000 \: {\rm K}$ & 15 \\
 & & $\phantom{k_{17}} = 9.6 \times 10^{-7} T^{-0.90}$ & $T > 8000 \: {\rm K}$ & \\
& & & & \\
18 & $\mH + \Dp \rightarrow \mD + \Hp$ & $k_{18} = 2.06 \times 10^{-10} T^{0.396}
\expf{-}{33}{T}$ & & 16 \\
& & $\phantom{k_{18}=} \mbox{} + 2.03 \times 10^{-9} T^{-0.332}$ & & \\
& & & & \\
19 & $\mD + \Hp \rightarrow \mH + \Dp$ & 
$k_{19} = 2.0 \times 10^{-10} T^{0.402} \expf{-}{37.1}{T}$ & $T \le 2 \times 10^{5} \: {\rm K}$ & 16 \\
& & $\phantom{k_{19} = } \mbox{} - 3.31 \times 10^{-17} T^{1.48}$ &  & \\
& & $\phantom{k_{19}} = 3.44 \times 10^{-10} T^{0.35}$ & $T > 2 \times 10^{5} \: {\rm K}$ & \\
& & & & \\
20 & $\mHt + \Dp \rightarrow \hd + \Hp$ &
$k_{20} = \left[0.417 + 0.846 \log{T} - 0.137 (\log{T})^{2} \right] \times 10^{-9}$ & & 17 \\
& & & & \\
21 & $\hd + \Hp \rightarrow \mHt + \Dp$ & $k_{21} = 1.1 \times 10^{-9} \expf{-}{488}{T}$ & & 17 \\
& & & & \\
22 & $\mHt + \mD \rightarrow \hd + \mH$ & 
$k_{22} = 1.69 \times 10^{-10} \expf{-}{4680}{T}$ & $T \leq 200 \: {\rm K}$ & 18 \\
& & $\phantom{k_{22}} =1.69 \times 10^{-10} \exp\left(-\frac{4680}{T} + 
\frac{198800}{T^{2}}\right)$ & $T > 200 \: {\rm K}$ & \\
& & & & \\
23 & $\hd + \mH \rightarrow \mD + \mHt$ & 
$k_{23} = 5.25 \times 10^{-11} \expf{-}{4430}{T}$ & $T \leq 200 \: {\rm K}$ & 19 \\
& & $\phantom{k_{23}} = 5.25 \times 10^{-11} \exp\left(-\frac{4430}{T} + \frac{173900}{T^{2}}\right)$
& $T > 200 \: {\rm K}$ & \\
& & & & \\
24 & $\He + \me \rightarrow \Hep + \me + \me$ &
$k_{24} =  \exp[-4.409864886 \times 10^{1}$ & & 13 \\
& & $\phantom{k_{24} = } \mbox{}  + 2.391596563 \times 10^{1} \ln T_{\rm e}$ & & \\
& & $\phantom{k_{24} = } \mbox{}  - 1.07532302 \times 10^{1} (\ln T_{\rm e})^{2}$ & & \\
& & $\phantom{k_{24} = } \mbox{}  + 3.05803875 \times 10^{0} (\ln T_{\rm e})^{3}$ & & \\
& & $\phantom{k_{24} = } \mbox{}  -  5.6851189 \times 10^{-1} (\ln T_{\rm e})^{4}$ & & \\
& & $\phantom{k_{24} = } \mbox{}  + 6.79539123 \times 10^{-2} (\ln T_{\rm e})^{5}$ & & \\
& & $\phantom{k_{24} = } \mbox{}   - 5.0090561 \times 10^{-3} (\ln T_{\rm e})^{6}$ & & \\ 
& & $\phantom{k_{24} = } \mbox{}  + 2.06723616\times 10^{-4} (\ln T_{\rm e})^{7}$ & & \\
& & $\phantom{k_{24} = } \mbox{}   -  3.64916141 \times 10^{-6} (\ln T_{\rm e})^{8}]$ & & \\
& & & & \\
25 & $\Hep + \me \rightarrow \He + \gamma$ & 
$k_{25, {\rm rr, A}} = 10^{-11} T^{-0.5} \left[12.72 - 1.615 \log{T} \right. $ & Case A & 20  \\
& & $\left. \phantom{k_{25, {\rm rr, A}} = } \mbox{} - 0.3162 (\log{T})^{2} + 0.0493 (\log{T})^{3}\right]$ & & \\
& & & & \\
& & $k_{25, {\rm rr, B}} = 10^{-11} T^{-0.5} \left[11.19 - 1.676 \log{T} \right. $ & Case B & 20  \\
& & $\left. \phantom{k_{25, {\rm rr, A}} = } \mbox{} - 0.2852 (\log{T})^{2} + 0.04433 (\log{T})^{3} \right]$ & & \\
& & & & \\
& & $k_{25, {\rm di}} = 1.9 \times 10^{-3} T^{-1.5} \expf{-}{473421}{T}$ & & \\
& & $\phantom{k_{25, {\rm di}} = } \mbox{} \times \left[1.0 + 0.3 \expf{-}{94684}{T} \right] $ & & 21 \\
& & & & \\
26 & $\Hep + \mH \rightarrow \He + \Hp$ & 
$k_{26} = 1.25 \times 10^{-15} \left(\frac{T}{300}\right)^{0.25}$ & & 22 \\
& & & & \\
27 & $\He + \Hp \rightarrow \Hep + \mH$ &
$k_{27} = 1.26 \times 10^{-9} T^{-0.75} \expf{-}{127500}{T}$ & $ T \leq 10000 \: {\rm K}$ & 23 \\
& & $\phantom{k_{27}} = 4.0 \times 10^{-37} T^{4.74}$ & $T > 10000 \: {\rm K}$ &  \\
& & & & \\
28 & $\Hep + \mD \rightarrow \He + \Dp$ & $k_{28} = k_{26}$ & & --- \\
& & & & \\
29 & $\He + \Dp \rightarrow \Hep + \mD$ &  $k_{29} = k_{27}$ & & --- \\
& & & & \\
30 & $\Cp + \me \rightarrow \mC  + \gamma$ &
$k_{30} = 4.67 \times 10^{-12}  \left(\frac{T}{300}\right)^{-0.6}$ & $T \le 7950 \: {\rm K}$ & 24 \\
& & $\phantom{k_{30} } =1.23 \times 10^{-17}  \left(\frac{T}{300}\right)^{2.49} 
\exp \left(\frac{21845.6}{T} \right)$ & $ 7950 \: {\rm K} < T \le 21140 \: {\rm K}$ & \\
& & $\phantom{k_{30}} = 9.62 \times 10^{-8} \left(\frac{T}{300}\right)^{-1.37} 
 \exp \left(\frac{-115786.2}{T} \right)$ & $T > 21140 \: {\rm K}$ & \\
 & & & & \\
31 & $\sip + \me \rightarrow \mSi + \gamma$ &
$k_{31} =  7.5 \times 10^{-12} \left(\frac{T}{300}\right)^{-0.55}$  & $T \le 2000 \: {\rm K}$ & 25 \\
& & $\phantom{k_{31}}= 4.86 \times 10^{-12} \left(\frac{T}{300}\right)^{-0.32}$ & 
$2000 \: {\rm K} < T \le 10^{4} \: {\rm K}$ & \\
& & $\phantom{k_{31}}= 9.08 \times 10^{-14} \left(\frac{T}{300}\right)^{0.818}$ & 
$T > 10^{4} \: {\rm K}$ & \\
 & & & & \\
 32 & $\op + \me  \rightarrow \mO + \gamma$ &
$k_{32} = 1.30 \times 10^{-10} T^{-0.64}$ &  $T \le 400 \: {\rm K}$ & 26 \\
& & $\phantom{k_{32}} = 1.41 \times 10^{-10} T^{-0.66} + 7.4 \times 10^{-4}  T^{-1.5}$ & & \\
& & $\phantom{k_{32}=} \mbox{} \times \exp \left(-\frac{175000}{T}\right) [1.0 + 0.062 \times 
\exp \left(-\frac{145000}{T}\right) ]$ & $T > 400 \: {\rm K}$ & \\
 & & & & \\
33 & $\mC  + \me  \rightarrow \Cp  + \me + \me$ & 
$k_{33} = 6.85 \times 10^{-8} (0.193 + u)^{-1} u^{0.25} e^{-u}$ & $u = 11.26 / T_{\rm e}$ & 27 \\
 & & & & \\
34 & $\mSi + \me  \rightarrow \sip + \me + \me$ & 
$k_{34} = 1.88 \times 10^{-7} (1.0 + u^{0.5}) (0.376 + u)^{-1} u^{0.25} e^{-u}$ & 
$ u = 8.2 / T_{\rm e}$ & 27 \\
 & & & & \\
35 & $\mO  + \me  \rightarrow \op  + \me + \me$ &
$k_{35} = 3.59 \times 10^{-8} (0.073 + u)^{-1} u^{0.34} e^{-u}$ & $u = 13.6 / T_{\rm e}$ & 27 \\
 & & & & \\
36 & $\op  + \mH  \rightarrow \mO  + \Hp$ &
$ k_{36} = 4.99 \times 10^{-11} T^{0.405} +
7.54 \times 10^{-10} T^{-0.458} $ & & 28 \\
 & & & & \\
37 & $\mO  + \Hp  \rightarrow \op  + \mH$ &
$k_{37} = [1.08 \times 10^{-11} T^{0.517} $ & & 29 \\
& & $\phantom{k_{37} = } \mbox{} + 4.00 \times 10^{-10} T^{0.00669}] \exp 
\left(-\frac{227}{T}\right)$ & & \\
 & & & & \\
 38 & $\mO + \Hep \rightarrow \op + \He$ & 
 $k_{38} = 4.991 \times 10^{-15} \left(\frac{T}{10000}\right)^{0.3794} 
 \expf{-}{T}{1121000}$ & & 30 \\
 & & $\phantom{k_{38} = } \mbox{} + 2.780 \times 10^{-15} 
 \left(\frac{T}{10000}\right)^{-0.2163} \expf{}{T}{815800}$ & & \\
 & & & & \\
39 & $\mC  + \Hp  \rightarrow \Cp  + \mH$ & $k_{39} = 3.9 \times 10^{-16} T^{0.213}$ & & 29 \\
 & & & & \\
40 & $\Cp  + \mH  \rightarrow \mC  + \Hp$ & $k_{40} = 6.08 \times 10^{-14} 
\left(\frac{T}{10000}\right)^{1.96} \expf{-}{170000}{T}$ & & 29 \\
 & & & & \\
 41 & $\mC + \Hep \rightarrow \cp + \He$ & 
 $k_{41} = 8.58 \times 10^{-17}  T^{0.757}$ & $T \leq 200 \: {\rm K}$ & 31 \\
 & & $\phantom{k_{41}} = 3.25 \times 10^{-17} T^{0.968}$ & $200 < T \leq 2000 \: {\rm K}$ & \\
 & & $\phantom{k_{41}} = 2.77 \times 10^{-19} T^{1.597}$ & $T > 2000 \: {\rm K}$ & \\
 & & & & \\ 
42 & $\mSi + \Hp  \rightarrow \sip + \mH$ &
$k_{42} = 5.88 \times 10^{-13} T^{0.848}$ & $T \le 10^{4} \: {\rm K}$ & 32 \\
& & $\phantom{k_{42}} = 1.45 \times 10^{-13} T$ & $T > 10^{4} \: {\rm K}$ & \\
 & & & & \\
 43 & $\mSi + \Hep \rightarrow \sip + \He$ & $k_{43} = 3.3 \times 10^{-9}$ & & 33 \\
 & & & & \\
44 & $\cp  + \mSi \rightarrow \mC + \sip$ & $k_{44} = 2.1 \times 10^{-9}$ & & 33 \\
 & & & & \\
 45 & $\sip + \Hp \rightarrow \sipp + \mH$ & 
 $k_{45} = 4.10 \times 10^{-10} \left( \frac{T}{10000} \right)^{0.24}$ & & 32 \\
 & & $\phantom{k_{45} = } \mbox{} \times \left[1.0 + 3.17 \expf{}{T}{2.39 \times 10^{6}} \right] 
 \expf{-}{3.178}{T_{\rm e}}$ & & \\
  & & & & \\
 46 & $\sipp + \mH \rightarrow \sip + \Hp$ & $k_{46} = 1.23 \times 10^{-9} 
 \left(\frac{T}{10000}\right)^{0.24}$ & & 32 \\ 
 & & $\phantom{k_{46} =} \mbox{} \times \left[1.0 + 3.17 \expf{}{T}{2.39 \times 10^{6}} \right] $ & & \\ 
  & & & & \\
 47 & $\sipp + \me \rightarrow \sip + \gamma$ & $k_{47, {\rm rr}} = 
 1.75 \times 10^{-12} \left( \frac{T}{10000} \right)^{-0.6346}$ & & 34 \\ 
 & & & & \\
 & & $k_{47, {\rm di}} = 2.2552 \times 10^{-11} T_{\rm e}^{-1.5} \expf{-}{2.76}{T_{\rm e}}$ & & 35 \\
& & $\phantom{k_{47, {\rm di}} = } \mbox{} + 5.6058 \times 10^{-9} T_{\rm e}^{-1.5} 
\expf{-}{10.13}{T_{\rm e}}$ & & \\
& & & & \\ 
75 & $\mHt + \He \rightarrow \mH + \mH + \He$ &
$k_{\rm 75, l} = {\rm dex}\left[\mbox{}-27.029 + 3.801 \log{(T)} - 29487/T \right]$ & & 36 \\ 
& & & & \\
& & $k_{\rm 75, h} = {\rm dex}\left[-2.729-1.75 \log{(T)} - 23474/T\right]$ & & \\
& & & & \\
& & $n_{\rm cr, He} = {\rm dex} \left[5.0792 (1.0 - 1.23 \times 10^{-5} (T - 2000) \right]$ & & 36 \\
& & & & \\
76 & $\oh + \mH \rightarrow \mO + \mH + \mH $ & $k_{76} = 6.0 \times 10^{-9} \expf{-}{50900}{T}$ & & 33 \\
& & & & \\
77 & ${\rm HOC^{+}} + \mHt \rightarrow {\rm HCO^{+}} + \mHt $ & $k_{77} = 3.8 \times 10^{-10}$ & & 37 \\
& & & & \\
78 & ${\rm HOC^{+}} + \co \rightarrow {\rm HCO^{+}} + \co $ & $k_{78} = 4.0 \times 10^{-10}$ & & 38 \\
& & & & \\
79 & $\mC + \mHt \rightarrow \ch + \mH $ &
$k_{79} = 6.64 \times 10^{-10} \expf{-}{11700}{T}$ & & 39 \\
& & & & \\
80 & $\ch + \mH \rightarrow \mC + \mHt$ & $k_{80} = 1.31 \times 10^{-10} \expf{-}{80}{T}$ & & 40 \\
& & & & \\
81 & $\ch + \mHt \rightarrow  \ch_{2} + \mH$ & 
$k_{81} = 5.46 \times 10^{-10} \expf{-}{1943}{T}$ & & 41 \\
& & & & \\
82 & $\ch + \mC \rightarrow \mC_{2} + \mH $ &
$k_{82} = 6.59 \times 10^{-11}$ & & 42 \\
& & & & \\
83 & $\ch + \mO \rightarrow \co + \mH $ & $k_{83} = 6.6 \times 10^{-11}$ & $T \le 2000 \: {\rm K}$  & 43 \\
& & $\phantom{k_{83}} = 1.02 \times 10^{-10} \expf{-}{914}{T}$ & $T > 2000 \: {\rm K}$ & 44 \\
& & & & \\
84 & $\ch_{2} + \mH \rightarrow \ch + \mHt$ & $k_{84} = 6.64 \times 10^{-11}$ & & 45 \\
& & & & \\
85 & $\ch_{2} + \mO \rightarrow \co + \mH + \mH$ & $k_{85} =1.33 \times 10^{-10}$ & & 46 \\
& & & & \\
86 & $\ch_{2} + \mO \rightarrow \co + \mHt$ & $k_{86} = 8.0 \times 10^{-11}$ & & 47 \\
& & & & \\
87 & $\mC_{2} + \mO \rightarrow \co + \mC $ & $k_{87} = 5.0 \times 10^{-11} \tmpt{0.5}$ & 
$T \le 300 \: {\rm K}$ & 48 \\
& & $\phantom{k_{87}} = 5.0 \times 10^{-11} \tmpt{0.757}$ & $T > 300 \: {\rm K}$ & 49 \\
& & & & \\
88 & $\mO + \mHt \rightarrow \oh + \mH$ & 
$k_{88} = 3.14 \times 10^{-13} \tmpt{2.7} \expf{-}{3150}{T}$ & & 50 \\
& & & & \\
89 & $\oh + \mH \rightarrow \mO + \mHt $ & 
$k_{89} = 6.99 \times 10^{-14} \tmpt{2.8} \expf{-}{1950}{T}$ & & 51 \\
& & & & \\
90 & $\oh + \mHt \rightarrow \hto + \mH$ & 
$k_{90} = 2.05 \times 10^{-12} \tmpt{1.52} \expf{-}{1736}{T}$ & & 52 \\
& & & & \\
91 & $\oh + \mC \rightarrow \co + \mH$ & $k_{91} = 1.0 \times 10^{-10}$ & & 42 \\
& & & & \\
92 & $\oh + \mO \rightarrow \mO_{2} + \mH $ &
$k_{92} = 3.50 \times 10^{-11}$  & $T \le 261 \: {\rm K}$ & 53 \\ 
& & $\phantom{k_{92}} = 1.77 \times 10^{-11} \expf{}{178}{T}$ & $T > 261 \: {\rm K}$ & 41 \\
& & & & \\
93 & $\oh + \oh \rightarrow \hto + \mH$ & 
$k_{93} = 1.65 \times 10^{-12} \tmpt{1.14} \expf{-}{50}{T}$ &  & 42 \\ 
& & & & \\
94 & $\hto + \mH \rightarrow \mHt + \oh $ &
$k_{94} =1.59 \times 10^{-11} \tmpt{1.2} \expf{-}{9610}{T}$ & & 54 \\
& & & & \\
95 & $\mO_{2} + \mH \rightarrow \oh + \mO $ & $k_{95} = 2.61 \times 10^{-10} \expf{-}{8156}{T}$ & & 41 \\
& & & & \\
96 & $\mO_{2} + \mHt \rightarrow \oh + \oh $ & $k_{96} = 3.16 \times 10^{-10} 
\expf{-}{21890}{T}$ & & 55 \\
& & & & \\
97 & $\mO_{2} + \mC \rightarrow \co + \mO $ &
$k_{97} = 4.7 \times 10^{-11} \tmpt{-0.34}$ & $T \le 295 \: {\rm K}$ & 42 \\
& & $ \phantom{k_{97}} = 2.48 \times 10^{-12} \tmpt{1.54} \expf{}{613}{T}$ & $T > 295 \: {\rm K}$ & 41 \\
& & & & \\
98 & $\co + \mH \rightarrow \mC + \oh $ & 
$k_{98} = 1.1 \times 10^{-10} \tmpt{0.5} \expf{-}{77700}{T}$ & & 33 \\
& & & & \\
99 & $\mHtp + \mHt \rightarrow \htp + \mH$ &
$k_{99} = 2.24 \times 10^{-9} \tmpt{0.042} \expf{-}{T}{46600}$ & & 56 \\
& & & & \\
100 & $\htp + \mH \rightarrow \mHtp + \mHt$ & $k_{100} = 7.7 \times 10^{-9} \expf{-}{17560}{T}$ & & 57 \\
& & & & \\
101 & $\mC + \mHtp \rightarrow \ch^{+} + \mH$ & $k_{101} = 2.4 \times 10^{-9}$ & & 33 \\
& & & & \\
102 & $\mC + \htp \rightarrow \ch^{+} + \mHt$ & $k_{102} = 2.0 \times 10^{-9}$ & & 33 \\
& & & & \\
103 & $\Cp + \mHt \rightarrow \ch^{+}  + \mH $ & 
$k_{103} = 1.0 \times 10^{-10} \expf{-}{4640}{T}$ & & 58 \\
& & & & \\
104 & $\ch^{+} + \mH \rightarrow \Cp + \mHt $ & $k_{104} = 7.5 \times 10^{-10}$ & & 59 \\
& & & & \\
105 & $\ch^{+} + \mHt \rightarrow \ch_{2}^{+} + \mH $ & $k_{105} = 1.2 \times 10^{-9}$ & & 59 \\
& & & & \\
106 & $\ch^{+} + \mO \rightarrow \co^{+} + \mH $ & $k_{106} = 3.5 \times 10^{-10}$ & & 60 \\
& & & & \\
107 & $\ch_{2} + \Hp \rightarrow \ch^{+} + \mHt $ & $k_{107} = 1.4 \times 10^{-9}$ & & 33 \\
& & & & \\
108 & $\ch_{2}^{+} + \mH \rightarrow \ch^{+} + \mHt $ & 
$k_{108} = 1.0 \times 10^{-9} \expf{-}{7080}{T}$ & & 33 \\
& & & & \\
109 & $\ch_{2}^{+} + \mHt \rightarrow \ch_{3}^{+} + \mH $ & $k_{109} = 1.6 \times 10^{-9}$ & & 61 \\
& & & & \\
110 & $\ch_{2}^{+} + \mO \rightarrow {\rm HCO}^{+} + \mH $ & $k_{110} = 7.5 \times 10^{-10}$ & & 33 \\
& & & & \\
111 & $\ch_{3}^{+} + \mH \rightarrow \ch_{2}^{+} + \mHt$ & 
$k_{111} = 7.0 \times 10^{-10} \expf{-}{10560}{T}$ & & 33 \\
& & & & \\
112 & $\ch_{3}^{+} + \mO \rightarrow {\rm HCO}^{+} + \mHt $ & $ k_{112} = 4.0 \times 10^{-10}$ & & 62 \\
& & & & \\
113 & $\mC_{2} + \Op \rightarrow \co^{+} + \mC $ & $k_{113} = 4.8 \times 10^{-10}$ & & 33 \\
& & & & \\
114 & $\Op + \mHt   \rightarrow  \oh^{+} + \mH $  & $k_{114} = 1.7 \times 10^{-9}$ & & 63 \\
& & & & \\
115 & $\mO + \mHtp  \rightarrow  \oh^{+} + \mH $ & $k_{115} = 1.5 \times 10^{-9}$ & & 33 \\
& & & & \\
116 & $\mO + \htp  \rightarrow  \oh^{+} + \mHt $  & $k_{116} = 8.4 \times 10^{-10}$ & & 64 \\
& & & & \\
117 & $\oh + \htp \rightarrow \hto^{+} + \mHt$ & $k_{117} = 1.3 \times 10^{-9}$ & & 33 \\
& & & & \\
118 & $\oh + \Cp \rightarrow \co^{+} + \mH $ & $k_{118} = 7.7 \times 10^{-10}$ & & 33 \\
& & & & \\
119 & $\oh^{+} + \mHt  \rightarrow  \hto^{+} + \mH $  & $k_{119} = 1.01 \times 10^{-9}$ & & 65 \\
& & & & \\
120 & $\hto^{+} + \mHt  \rightarrow  {\rm H_{3}O}^{+} + \mH $  & 
$k_{120} = 6.4 \times 10^{-10}$ & & 66 \\
& & & & \\
121 & $\hto + \htp \rightarrow {\rm H_{3}O^{+}} + \mHt$ &$k_{121} = 5.9 \times 10^{-9}$ & & 67 \\
& & & & \\
122 & $\hto + \Cp \rightarrow {\rm HCO^{+}} + \mH$ & $k_{122} = 9.0 \times 10^{-10}$ & & 68 \\
& & & & \\
123 & $\hto +\Cp \rightarrow {\rm HOC^{+}} + \mH$ & $k_{123} = 1.8 \times 10^{-9}$ & & 68 \\
& & & & \\
124 & ${\rm H_{3}O}^{+} + \mC  \rightarrow {\rm HCO^{+}} + \mHt$ & $k_{124} = 1.0 \times 10^{-11}$  
&  & 33 \\
& & & & \\
125 & ${\rm H_{3}O}^{+} + \msi  \rightarrow \hto + {\rm SiH^{+}} $  & 
$k_{125} = 1.8 \times 10^{-9}$ & & 33 \\
& & & & \\
126 & $\mO_{2} + \Cp \rightarrow \co^{+} + \mO $ & $k_{126} = 3.8 \times 10^{-10}$ & & 61 \\
& & & & \\
127 & $\mO_{2} + \Cp \rightarrow \co + \Op $ & $k_{127} = 6.2 \times 10^{-10}$ & & 61 \\
& & & & \\
128 & $\mO_{2} + \ch_{2}^{+} \rightarrow {\rm HCO}^{+} + \oh $ & 
$k_{128} = 9.1 \times 10^{-10}$ & & 61 \\
& & & & \\
129 & $\mO_{2}^{+} + \mC \rightarrow \co^{+} + \mO $ & $k_{129} = 5.2 \times 10^{-11}$ & & 33 \\
& & & & \\
130 & $\co + \htp \rightarrow {\rm HOC^{+}} + \mHt$ & $k_{130} = 2.7 \times 10^{-11}$ & & 69 \\
& & & & \\
131 & $\co + \htp \rightarrow {\rm HCO^{+}} + \mHt$ & $k_{131} = 1.7 \times 10^{-9}$ & & 69 \\
& & & & \\
132 & ${\rm HCO^{+}} + \mC \rightarrow \co + \ch^{+}$ & $k_{132} = 1.1 \times 10^{-9}$ & & 33 \\
& & & & \\
133 & ${\rm HCO^{+}} + \msi \rightarrow \co + {\rm SiH^{+}}$ & $k_{133} = 1.6 \times 10^{-9}$ & & 33 \\
& & & & \\
134 & ${\rm HCO^{+}} + \hto \rightarrow \co + {\rm H_{3}O^{+}} $ & 
$k_{134} = 2.5 \times 10^{-9}$ & & 70 \\
& & & & \\
135 & $\msi + \htp \rightarrow {\rm SiH^{+}} + \mHt$ & $k_{135} = 3.7 \times 10^{-9}$ & & 33 \\
& & & & \\
136 & ${\rm SiH^{+}} + \mH  \rightarrow  \sip  + \mHt $ & $k_{136} = 1.9 \times 10^{-9}$ & & 71 \\
& & & & \\
137 & ${\rm SiH^{+}} + \hto \rightarrow \msi + {\rm H_{3}O^{+}}$ & 
$k_{137} = 8.0 \times 10^{-10}$ & & 71 \\
& & & & \\
138 & $\mHt + \Hep \rightarrow \He + \mHtp$ & $k_{138} = 7.2 \times 10^{-15}$ & & 72 \\
& & & & \\
139 & $\mHt  + \Hep \rightarrow \He + \mH + \Hp$ & $k_{139} = 3.7 \times 10^{-14} \expf{}{35}{T}$ 
& & 72 \\  
& & & & \\
140 & $\ch + \Hp \rightarrow \ch^{+} + \mH$ & $k_{140} = 1.9 \times 10^{-9}$ & & 33 \\
& & & & \\
141 & $\ch_{2} + \Hp \rightarrow \ch_{2}^{+} + \mH$ & $k_{141} = 1.4 \times 10^{-9}$ & & 33 \\
& & & & \\
142 & $\ch_{2} + \Hep \rightarrow \Cp + \He + \mHt$ & $k_{142} = 7.5 \times 10^{-10}$ & & 33 \\
& & & & \\ 
143 & $\mC_{2} + \Hep \rightarrow \Cp + \mC + \He $ & $k_{143} =1.6 \times 10^{-9}$  & & 33 \\
& & & & \\
144 & $\oh + \Hp \rightarrow \oh^{+} + \mH$ & $k_{144} = 2.1 \times 10^{-9}$ & & 33 \\
& & & & \\
145 & $\oh + \Hep \rightarrow \Op + \He + \mH$ & $k_{145} = 1.1 \times 10^{-9}$ & & 33 \\
& & & & \\
146 & $\hto + \Hp \rightarrow \hto^{+} + \mH $ & $k_{146} = 6.9 \times 10^{-9}$ & & 73 \\
& & & & \\
147 & $\hto + \Hep \rightarrow \oh + \He + \Hp $ & $k_{147} = 2.04 \times 10^{-10}$ & & 74 \\
& & & & \\
148 & $\hto + \Hep \rightarrow \oh^{+} + \He + \mH $ & $k_{148} = 2.86 \times 10^{-10}$ & & 74 \\
& & & & \\
149 & $\hto + \Hep \rightarrow \hto^{+} + \He $ & $k_{149} = 6.05 \times 10^{-11}$ & & 74 \\
& & & & \\
150 & $\mO_{2} + \Hp  \rightarrow \mO_{2}^{+} + \mH $ & $k_{150} = 2.0 \times 10^{-9}$ & & 73 \\
& & & & \\
151 & $\mO_{2} + \Hep \rightarrow \mO_{2}^{+} + \He $ & $k_{151} = 3.3 \times 10^{-11}$ & & 75 \\
& & & & \\
152 & $\mO_{2} + \Hep \rightarrow \Op + \mO + \He $ & $k_{152} = 1.1 \times 10^{-9}$ & & 75 \\
& & & & \\
153 & $\mO_{2}^{+} + \mC \rightarrow \mO_{2} + \Cp $ & $k_{153} = 5.2 \times 10^{-11}$ & & 33 \\
& & & & \\
154 & $\co + \Hep \rightarrow \Cp + \mO + \He $ & $k_{154} = 1.4 \times 10^{-9} \tmpt{-0.5}$ & & 76 \\
& & & & \\
155 & $\co + \Hep \rightarrow \mC + \Op + \He $ & $k_{155} = 1.4 \times 10^{-16} \tmpt{-0.5}$ & & 76 \\
& & & & \\
156 & $\co^{+} + \mH \rightarrow \co + \Hp $ & $k_{156} = 7.5 \times 10^{-10}$ & & 77 \\
& & & & \\
157 & $\Cm + \Hp \rightarrow \mC + \mH $ & $k_{157} = 2.3 \times 10^{-7} \tmpt{-0.5}$ & & 33 \\
& & & & \\
158 & $\Om + \Hp \rightarrow \mO + \mH $ & 
$k_{158} = 2.3 \times 10^{-7} \tmpt{-0.5}$ & & 33 \\ 
& & & & \\
159 & $\Hep + \Hm \rightarrow \He + \mH$ & $k_{159} = 2.32 \times 10^{-7} \tmpt{-0.52} 
\expf{}{T}{22400}$ & & 78 \\
& & & & \\
160 & $\htp + \me \rightarrow \mHt + \mH$ & $k_{160} = 2.34 \times 10^{-8} \tmpt{-0.52}$ & & 79 \\
& & & & \\
161 & $\htp + \me \rightarrow \mH + \mH + \mH$ & $k_{161} = 4.36 \times 10^{-8} \tmpt{-0.52}$ & & 79 \\
& & & & \\ 
162 & $\ch^{+} + \me \rightarrow \mC + \mH $ & $k_{162} = 7.0 \times 10^{-8} \tmpt{-0.5}$ & & 80 \\
& & & & \\
163 & $\ch_{2}^{+} + \me \rightarrow \ch + \mH$ & $k_{163} = 1.6 \times 10^{-7} \tmpt{-0.6}$  & & 81 \\
& & & & \\
164 & $\ch_{2}^{+} + \me \rightarrow \mC + \mH + \mH$ & 
$k_{164} = 4.03 \times 10^{-7} \tmpt{-0.6}$ & & 81 \\
& & & & \\
165 & $\ch_{2}^{+} + \me \rightarrow \mC + \mHt$ & $k_{165} = 7.68 \times 10^{-8} \tmpt{-0.6}$  & & 81 \\
& & & & \\
166 & $\ch_{3}^{+} + \me \rightarrow  \ch_{2} + \mH$ & 
$k_{166} = 7.75 \times 10^{-8} \tmpt{-0.5}$ & & 82 \\
& & & & \\
167 & $\ch_{3}^{+} + \me \rightarrow \ch + \mHt$ & $k_{167} = 1.95 \times 10^{-7} \tmpt{-0.5}$ & & 82 \\
& & & & \\
168 & $\ch_{3}^{+} + \me \rightarrow \ch + \mH + \mH$ & 
$k_{168} = 2.0 \times 10^{-7} \tmpt{-0.4}$ & & 33 \\
& & & & \\
169 & $\oh^{+} + \me \rightarrow \mO + \mH$ & $k_{169} = 6.3 \times 10^{-9} \tmpt{-0.48}$ & & 83 \\
& & & & \\
170 & $\hto^{+} + \me \rightarrow \mO + \mH + \mH$ & 
$k_{170} = 3.05 \times 10^{-7} \tmpt{-0.5}$ & & 84 \\
& & & & \\
171 & $\hto^{+} + \me \rightarrow \mO + \mHt $ & $k_{171} = 3.9 \times 10^{-8} \tmpt{-0.5}$  & & 84 \\
& & & & \\
172 & $\hto^{+} + \me \rightarrow \oh + \mH $ & $k_{172} = 8.6 \times 10^{-8} \tmpt{-0.5}$ & & 84 \\
& & & & \\
173 & ${\rm H_{3}O}^{+} + \me  \rightarrow \mH + \hto $  & 
$k_{173} = 1.08 \times 10^{-7} \tmpt{-0.5}$ & & 85 \\
& & & & \\
174 & ${\rm H_{3}O}^{+} + \me  \rightarrow \oh + \mHt $  & 
$k_{174} = 6.02 \times 10^{-8} \tmpt{-0.5}$ & & 85 \\
& & & & \\
175 & ${\rm H_{3}O}^{+} + \me  \rightarrow \oh + \mH + \mH $  & 
$k_{175} = 2.58 \times 10^{-7} \tmpt{-0.5}$ & & 85 \\
& & & & \\
176 & ${\rm H_{3}O}^{+} + \me  \rightarrow \mO + \mH + \mHt $  & 
$k_{176} = 5.6 \times 10^{-9}  \tmpt{-0.5}$ & & 85 \\
& & & & \\
177 & $\mO_{2}^{+} + \me \rightarrow \mO + \mO $ & $k_{177} = 1.95 \times 10^{-7} \tmpt{-0.7}$ & & 86 \\
& & & & \\
178 & $\co^{+} + \me \rightarrow \mC + \mO$ & $k_{178} = 2.75 \times 10^{-7} \tmpt{-0.55}$ & & 87 \\
& & & & \\ 
179 & ${\rm HCO^{+}} + \me \rightarrow \co + \mH $ & 
$k_{179} = 2.76 \times 10^{-7} \tmpt{-0.64}$ & & 88 \\
& & & & \\
180 & ${\rm HCO^{+}} + \me \rightarrow \oh + \mC $ & 
$k_{180} = 2.4 \times 10^{-8} \tmpt{-0.64}$  & & 88 \\
& & & & \\
181 & ${\rm HOC^{+}} + \me \rightarrow \co + \mH $ & $k_{181} = 1.1 \times 10^{-7} \tmpt{-1.0}$ & & 33 \\ 
& & & & \\ 
182 & ${\rm SiH^{+}}  + \me  \rightarrow  \msi + \mH $ & $k_{182} = 2.0 \times 10^{-7} \tmpt{-0.5}$ & & 33 \\
& & & & \\
183 & $\Hm + \mC \rightarrow \ch + \me$ & $k_{183} = 1.0 \times 10^{-9}$ & & 33 \\
& & & & \\ 
184 & $\Hm + \mO \rightarrow \oh + \me$ & $k_{184} = 1.0 \times 10^{-9}$ & & 33 \\
& & & & \\ 
185 & $\Hm + \oh \rightarrow \hto + \me$ & $k_{185} = 1.0 \times 10^{-10}$ & & 33 \\
& & & & \\
186 & $\Cm + \mH \rightarrow \ch + \me $ & $k_{186} = 5.0 \times 10^{-10}$ & & 33 \\
& & & & \\
187 & $\Cm + \mHt \rightarrow \ch_{2} + \me $ & $k_{187} = 1.0 \times 10^{-13}$ & & 33 \\
& & & & \\
188 & $\Cm + \mO \rightarrow \co + \me $ & $k_{188} = 5.0 \times 10^{-10}$ & & 33 \\
& & & & \\
189 & $\Om + \mH \rightarrow \oh + \me$ & $k_{189} = 5.0 \times 10^{-10}$ & & 33 \\
& & & & \\
190 & $\Om + \mHt \rightarrow \hto + \me $ & $k_{190} = 7.0 \times 10^{-10}$ & & 33 \\
& & & & \\
191 & $\Om + \mC \rightarrow \co + \me $ & $k_{191} = 5.0 \times 10^{-10}$ & & 33 \\
& & & & \\
192 & $\mHt + \Hp \rightarrow \htp + \gamma$ & 
$k_{192} = 1.0 \times 10^{-16}$ & & 89 \\
& & & & \\ 
193 & $\mC + \me \rightarrow \Cm + \gamma $ & $k_{193} = 2.25 \times 10^{-15}$ & & 90 \\
& & & & \\
194 & $\mC + \mH \rightarrow \ch + \gamma$ & $k_{194} = 1.0 \times 10^{-17}$ & & 91 \\
& & & & \\
195 & $\mC + \mHt \rightarrow \ch_{2} + \gamma$ & $k_{195} = 1.0 \times 10^{-17}$ & & 91 \\
& & & & \\
196 & $\mC + \mC \rightarrow \mC_{2} + \gamma $ & 
$k_{196} = 4.36 \times 10^{-18} \tmpt{0.35} \expf{-}{161.3}{T}$ & & 92 \\
& & & & \\
197 & $\mC + \mO \rightarrow \co + \gamma$ & 
$k_{197} = 2.1 \times 10^{-19}$ & $T \le 300 \: {\rm K}$ & 93 \\
& & $\phantom{k_{197}} = 3.09 \times 10^{-17} \tmpt{0.33} \expf{-}{1629}{T}$ & 
$T > 300 \: {\rm K}$ & 94 \\
& & & & \\
198 & $\Cp + \mH  \rightarrow \ch^{+}  + \gamma $ & 
$k_{198} = 4.46 \times 10^{-16} T^{-0.5} \expf{-}{4.93}{T^{2/3}}$ & & 95 \\
& & & & \\
199 & $\Cp + \mHt \rightarrow \ch_{2}^{+} + \gamma $ & 
$k_{199} = 4.0 \times 10^{-16} \tmpt{-0.2}$ & & 96 \\
& & & & \\
200 & $\Cp + \mO \rightarrow \co^{+} + \gamma$ & $k_{200} = 2.5 \times 10^{-18}$ & 
$T  \le 300 \: {\rm K}$ & 93  \\
& & $\phantom{k_{200}} = 3.14 \times 10^{-18} \tmpt{-0.15} \expf{}{68}{T}$ & $T > 300 \: {\rm K}$ & \\
& & & & \\
201 & $\mO + \me \rightarrow \mO^{-} + \gamma$ & $k_{201} = 1.5 \times 10^{-15}$ & & 33 \\
& & & & \\
202 & $\mO + \mH \rightarrow \oh + \gamma$ & $k_{202} = 9.9 \times 10^{-19} \tmpt{-0.38}$ & & 33 \\
& & & & \\
203 & $\mO + \mO \rightarrow \mO_{2} + \gamma $ & 
$k_{203} = 4.9 \times 10^{-20} \tmpt{1.58}$ & & 91 \\
& & & & \\
204 & $\oh + \mH \rightarrow \hto + \gamma$ & 
$k_{204} = 5.26 \times 10^{-18} \tmpt{-5.22} \expf{-}{90}{T}$ & & 97 \\
& & & & \\
205 & $\sip  + \mH   \rightarrow {\rm SiH^{+}}  + \gamma $ & 
$k_{205} = 1.17 \times 10^{-17} \tmpt{-0.14}$ & & 98 \\
& & & & \\
206 & $\mH + \mH + \mH \rightarrow \mHt + \mH$ & 
$k_{206} = 1.32 \times 10^{-32} \tmpt{-0.38}$ & $T \le 300 \: {\rm K}$ & 99 \\ 
& & $\phantom{k_{206}} = 1.32 \times 10^{-32} \tmpt{-1.0}$ & $T > 300 \: {\rm K}$ & 100 \\ 
& & & & \\
207 & $\mH + \mH + \mHt \rightarrow \mHt + \mHt$ &
$k_{207} = 2.8 \times 10^{-31} T^{-0.6}$ & & 101 \\ 
& & & & \\
208 & $\mH + \mH + \He \rightarrow \mHt + \He$ & $k_{208} = 6.9 \times 10^{-32} T^{-0.4}$ & & 102 \\
& & & & \\
209 & $\mC + \mC + {\rm M} \rightarrow \mC_{2} + {\rm M}$ &
$k_{209} = 5.99 \times 10^{-33} \tmptscl{5000}{-1.6}$ & $T \le 5000 \: {\rm K}$ & 103 \\
& & $\phantom{k_{209}} = 5.99 \times 10^{-33} \tmptscl{5000}{-0.64} \expf{}{5255}{T}$ &
$T > 5000 \: {\rm K}$ & 104 \\ 
& & & & \\
210 & $\mC + \mO + {\rm M} \rightarrow \co + {\rm M}$ &
$k_{210} = 6.16 \times 10^{-29} \tmpt{-3.08}$& $T \le 2000 \: {\rm K}$ & 105 \\ 
& & $\phantom{k_{210}} = 2.14 \times 10^{-29} \tmpt{-3.08} \expf{}{2114}{T}$ & 
$T > 2000 \: {\rm K}$ & 76 \\
& & & & \\
211 & $\Cp + \mO + {\rm M} \rightarrow \co^{+} + {\rm M}$ & $k_{211} = 100 \times k_{210}$ & & 76 \\
& & & & \\
212 & $\mC + \Op + {\rm M} \rightarrow \co^{+} + {\rm M}$ & $k_{212} = 100 \times k_{210}$ & & 76 \\
& & & & \\
213 & $\mO + \mH + {\rm M} \rightarrow \oh+ {\rm M}$ & 
$k_{213} = 4.33 \times 10^{-32} \tmpt{-1.0}$ & & 51 \\
& & & & \\
214 & $\oh + \mH + {\rm M} \rightarrow \hto + {\rm M}$ & 
$k_{214} = 2.56 \times 10^{-31} \tmpt{-2.0}$ & & 105 \\
& & & & \\
215 & $\mO + \mO + {\rm M} \rightarrow \mO_{2} + {\rm M}$ &
$k_{215} = 9.2 \times 10^{-34} \tmpt{-1.0}$ & & 45 \\
& & & & \\
216 & $\mO + \ch \rightarrow {\rm HCO^{+}} + \me$ & 
$k_{216} = 2.0 \times 10^{-11} \tmpt{0.44}$ & & 106 \\
\enddata
\tablerefs{1: \citet{WIS79}, 2: \citet{LAU91}, 3: \citet{RAM76}, 4: \citet{KAR79},
5: \citet{MOS70}, 6: \citet{SCH94}, 7: \citet{SAV04}, 8: \citet{STI99}, 9: \citet{MAC86},
10: \citet{ls83}, 11: \citet{mkm98}, 12: \citet{sk87},
13: \citet{JAN87}, 14: \citet{FER92}, 15: \citet{POU78}, 
16: \citet{sav02}, 17: \citet{ger82}, 18: \citet{mlts94}, 19: \citet{s59}, 
20: \citet{hs98}, 21: \citet{ap73}, 22: \citet{z89}, 23: \citet{kldd93}, 
24: \citet{NAH97}, 25: \citet{NAH00}, 26: \citet{NAH99}, 27: \citet{VOR97},
28: \citet{STA99}, 29: \citet{STA98}, 30: \citet{z04}, 31: \citet{kd93},
32: \citet{KIN96}, 33: \citet{TEU00}, 34: \citet{n95,n96}, 35: \citet{mmcv98},
36: \citet{drcm87}, 37: \citet{ssrg02}, 38: \citet{wr85}, 39: \citet{ddh91},
40: \citet{hgs93}, 41: Fit by \citet{TEU00} to data from the NIST 
chemical kinetics database; original source or sources unclear,
42: \citet{shc04},  43: \citet{bau92}, 44 \citet{mr86},
45: \citet{war84}, 46: \citet{fra86}, 47: Fit by \citet{TEU00} to 
data from \citet{fj84} and \citet{fbj88}, 48: \citet{md83}, 
49: \citet{fa69} and paper III, 50: \citet{nr87}, 51: \citet{th86},
52: \citet{old92}, 53: \citet{cgk06} and paper III, 
54: \citet{cw79}, 55: \citet{aat75}, 56: \citet{ljb95}, 57: \citet{smt92},
58: \citet{asm84}, 59: \citet{ms99}, 60: \citet{vig80},
61: \citet{sa77a,sa77b}, 62: \citet{feh76}, 63: \citet{sam78,asp80},
64: \citet{mm00}, 65: \citet{jbt81}, 66: \citet{rw80}, 67: \citet{kth74,afhk75},
68: \citet{ahf76,wah76}, 69: \citet{kth75}, 70: \citet{asg78}, 
71: \citet{herb89}, 72: \citet{ba84}, 73: \citet{ssm92}, 
74: \citet{mdm78a,mdm78b}, 75: \citet{as76a, as76b}, 76: \citet{pd89}, 
77: \citet{fed84a,fed84b}, 78: \citet{ph94}
79: Fit by \citet{umist07} to data from \citet{mac04}, 80: \citet{tkl91},
81: \citet{lls98}, 82: \citet{mi90},  83: \citet{gub95}, 84: \citet{rds00}, 
85: \citet{jbs00}, 86: \citet{aas83}, 87: \citet{rpl98},  88: \citet{gep05},
89: \citet{gh92}, 90: \citet{sd98}, 91: \citet{ph80}, 92: \citet{as97}, 
93: \citet{dd90}, 94: \citet{ssbo99}, 95: \citet{bvh06}, 96: \citet{he85}, 
97: \citet{fa80}, 98: \citet{skg00}, 99: \citet{or87}, 100: \citet{abn02}
101: \citet{cw83},  102: \citet{wk75},  103: Paper III, 
104: Fit by \citet{TEU00} to data from \citet{fa69} and \citet{sla76},
105: \citet{bau92}, 106: \citet{mb73}}
\tablecomments{$T$ and $T_{\rm e}$ are the gas temperature in units of K 
and eV respectively. References are to the primary source of data for each 
reaction.}
\end{deluxetable}

\begin{deluxetable}{lc}
\tabletypesize{\small}
\tablecaption{List of processes in our thermal model. \label{tab:cool}}
\tablewidth{0pt}
\tablehead{
\colhead{Process}  & \colhead{References} }
\startdata
{\bf Cooling:} & \\
Lyman-$\alpha$ cooling & \citet{CEN92} \\
He electronic excitation & \citet{CEN92,BBFT00} \\
Thermal bremsstrahlung & \citet{sk87} \\
Compton cooling & \citet{CEN92} \\
$\mHt$ rotational, vibrational lines & \citet{BOU99} \\
$\hd$ rotational, vibrational lines  & \citet{lna05} \\
Fine structure lines ($\mC, \Cp, \mO, \msi, \sip$) & Many sources; see Paper I \\
CO rotational, vibrational lines & \citet{nk93,nlm95} \\
${\rm H_{2}O}$ rotational, vibrational lines & \citet{nk93,nlm95} \\
OH rotational, vibrational lines  lines & \citet{pav02} \\
$\Hp$ recombination & \citet{FER92} \\
$\Hep$ recombination & \citet{ap73,hs98} \\
$\mH$ \& $\He$ collisional ionization & \citet{JAN87} \\
$\mHt$ collisional dissociation & See \S\ref{tab:chem_gas} \\
{\bf Heating:} & \\
$\mHt$ gas-phase formation & Many sources; see Paper I and \S\ref{tab:chem_gas} \\
\enddata
\end{deluxetable}

\end{document}